\documentclass[english,iopart]{iopart}

\usepackage[T1]{fontenc}
\usepackage[latin9]{inputenc}
\usepackage{geometry}
\usepackage{array}
\usepackage{float}
\usepackage{textcomp}
\usepackage{multirow}
\usepackage{amsbsy}
\usepackage{amstext}
\usepackage{graphicx}
\usepackage[numbers]{natbib}

\makeatletter

\providecommand{\tabularnewline}{\\}
\floatstyle{ruled}
\newfloat{algorithm}{tbp}{loa}
\providecommand{\algorithmname}{Algorithm}
\floatname{algorithm}{\protect\algorithmname}

\usepackage{iopams}
\usepackage{setstack}

\newcommand{\eqref}[1]{(\ref{#1})}

\usepackage{algorithm}
\usepackage{algorithmic}
\usepackage{graphicx}
\usepackage{iopams}

\@ifundefined{showcaptionsetup}{}{%
 \PassOptionsToPackage{caption=false}{subfig}}
\usepackage{subfig}
\makeatother

\usepackage{babel}
\begin{document}
\title{Expectation-Maximization Algorithm for Identification of Mesh-based
Compartment Thermal Model of Power Modules}
\title{}
\author{{\Large{}J Šev\v{c}ík}\textsuperscript{1}{\Large{}, V Šmídl}\textsuperscript{2}{\Large{}
and O Straka}\textsuperscript{3}}
\address{\textsuperscript{1,2}{\large{}Research and Innovation Centre for
Electrical Engineering, University of West Bohemia, Pilsen, Czech
Republic}}
\address{\textsuperscript{3}{\large{}Faculty of Applied Sciences, University
of West Bohemia, Pilsen, Czech Republic}}
\ead{{\large{}jsevcik@rice.zcu.cz, vsmidl@rice.zcu.cz, straka30@kky.zcu.cz}
}
\begin{abstract}
Accurate knowledge of temperatures in power semiconductor modules
is crucial for proper thermal management of such devices. Precise
prediction of temperatures allows to operate the system at the physical
limit of the device avoiding undesirable over-temperatures and thus
improve reliability of the module.

Commonly used thermal models can be based on detailed expert knowledge
of the device's physical structure or on precise and complete temperature
distribution measurements. The latter approach is more often used
in the industry. Recently, we have proposed a linear time invariant
state-space thermal model based on a compartment representation and
its identification procedure that is based on the Expectation-Maximization
algorithm from incomplete temperature data. However, the model still
requires to measure temperatures of all active elements.

In this contribution, we aim to relax the need for all measurements.
Therefore, we replace the previous dark gray-box approach with a structured
compartment model. The structure of the model is designed by a mesh-based
discretization of the physical layout of the module. All compartments
are assumed to share parameters that are identified from the data
of the measured elements. Temperatures of the unmeasured elements
are predicted using the shared parameters.

Identification of the parameters is possible only with suitable regularization
due to limited amount of the data. In particular, the model tightening
is accomplished by sharing parameters among compartments and by constraining
the process covariance matrix of the model in this contribution. Applicability
of the proposed identification procedure is discussed in terms of
growing state-space and therefore speeding up of the identification
algorithm is suggested. Performance of the proposed approach is tested
and demonstrated on simulated data.
\end{abstract}
\maketitle

\section{Introduction}

Monitoring of temperature distribution and its accurate prediction
in power semiconductor modules is fundamental for the proper thermal
management, that enables to operate system at the physical limit and
prevents device failures due to undesirable thermal stresses. Therefore,
the integration of the precise thermal model into the thermal protection
algorithm is essential, since the direct measurement of all temperature
distribution is very often infeasible (e.g. due to necessity for device
capsuling or low cost production claims).

A popular class of models used for heat transfer simulations are models
based on numerical discretization such as Finite Difference Method
(FDM), Finite Element Method (FEM) or Finite Volume Method (FVM).
These methods yield very accurate results but at the expense of high
computational requirements and thus they cannot be used in online
prediction. Moreover, the design and especially validation of these
models can be very time-consuming and without precise knowledge of
device's physical structure difficultly realizable.

Another class of models is based on Lumped Parameter Thermal Networks
(LPTNs) using thermal resistors and thermal capacitors as an analoqy
to electrical circuits for modeling of the heat transfer in the devices.
Generally, LPTNs produce reasonably accurate results requiring much
less computational time in comparison to models based on numerical
discretization methods. LPTNs can be classified as dark gray-box,
light gray-box or white-box models depending on the number of used
equivalent Resistor and Capacitor (RC) elements \citep{SSmodel-LPTN-Wallscheid}.
Dark gray-box LPTNs use only units of elements and therefore computational
requirements of such models can be very low. On the contrary dark
gray-box LPTNs are strongly abstracted, the information about temperature
distribution in the device is not complete (only selected points in
the device can be monitored) and several thermal phenomena (e.g. coupling
effect or temperature distribution in the segment like a chip) are
ignored.

Increasing microcontrollers computational performance enables to employ
improved LPTNs with more complicated structure (light gray-box or
white-box models).  Using higher level of elements in the LPTN can
lead to improved solution accuracy \citep{Boglietti_modernApproachesForThermalAnalysis}
and finer details, e.g. spatial temperature distribution, boundary
conditions or coupling effect \citep{lateral_coupling_LPTN_review,fosterRC-Bahman}
can be covered in the model. Nevertheless, RC parameters of such models
are very often extracted from the complicatedly calibrated FEM model
using transient (step) response analysis and following exponential
fitting techniques applied on transient thermal impedance curves.
An interesting approach how to create light gray- or white-box RC
model is to uses a mesh-based LPTN \citep{mesh-based-actuator,mesh-based-bar},
that can be identify from a geometric and material description of
the device. This standard identification procedure is strongly dependent
on quality of information about device physical structure and still
requires rich experience to obtain reasonable results~\citep{Boglietti_modernApproachesForThermalAnalysis}.

Recently, authors of this contribution proposed the Linear Time Invariant
(LTI) State Space (SS) compartment thermal model and its self-tuning
identification using Expectation-Maximization (EM) algorithm in \citep{our_iecon2019}.
This approach enables identification of the model from incomplete
temperature data and allows to combine sets of various measurements
of Temperature Sensitive Electrical Parameters (TSEPs) or direct measurements.
However, the model still requires to measure temperatures of all active
elements. In this paper, we aim to relax the need for all measurements.
Therefore we investigate the mesh-based structured compartment model
in this paper. In the sense of classification into the black-/gray-/white-box
model, \citep{our_iecon2019} can be seen as dark gray-box model.
The proposed mesh-based model falls into white-box models, since there
is much finer compartment structure based on discretization of the
physical layout of the module. This results in a growing state space
of temperatures and growing dimension of the state matrix in the proposed
LTI SS model. For that reason, the model is tightened by sharing parameters,
that are identified from the data. The possibility to apply Expectation-Maximization
algorithm for the identification of the consequently growing LTI SS
is studied in detail.

\section{Mesh-based compartment structure}

The proposed LTI SS thermal model is based on a \textit{compartment}
representation of the studied power semiconductor module. The compartment
model may be understood as a coarsely discretized model in the sense
of numerical methods. In other words, each compartment stands for
a relatively spacious control volume of the area of interest. In comparison
to the numerical discretization, the module is possible to be represented
by only units of compartments, although a finer compartment representation
can give better results. For that reason, the models based on mesh
representation and comprising structured compartments (circa hundreds
of compartments what is still much less than it is usual in numerical
models) are objects of our interest.

The three-dimensional volume of entire domain of the investigated
power module is discretized into rectangular elements (cubes or cuboids)
in a uniform Cartesian grid. To these elements we refer as to compartments.
Each basic-sized compartment can be further refined into four finer
compartments in X- and Y-axis in quadtree sense in the case that the
compartment contains more various components (e.g. a part of diode's
volume and a part of transistor's volume are located in the same compartment).

\begin{figure}
\begin{centering}
\begin{minipage}[t]{0.65\columnwidth}%
\begin{center}
\subfloat[Scheme of the layer arrangement\label{fig:layerArrangment}]{\begin{centering}
\vphantom{\includegraphics[width=0.523\columnwidth]{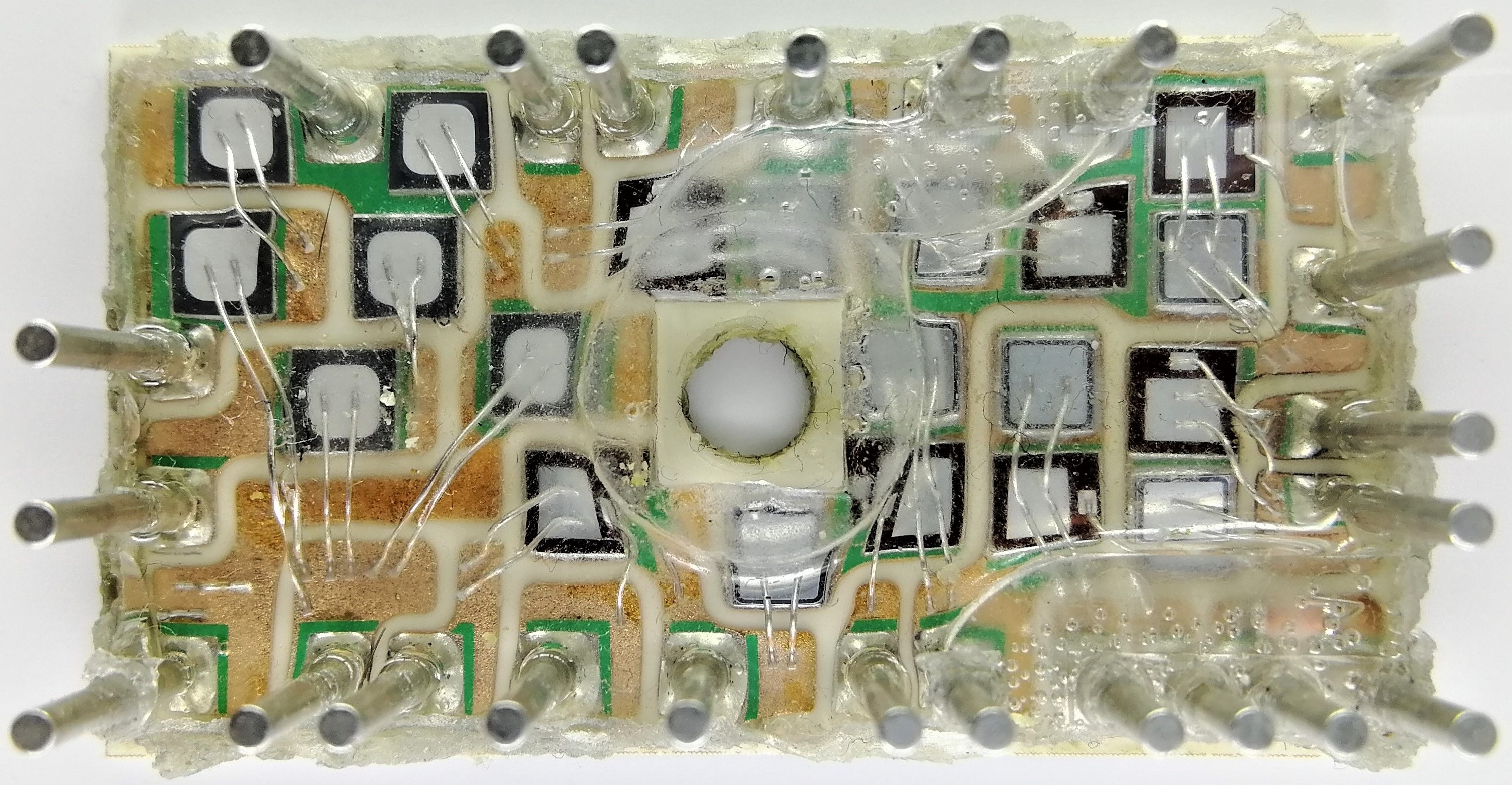}}\includegraphics[width=1\columnwidth]{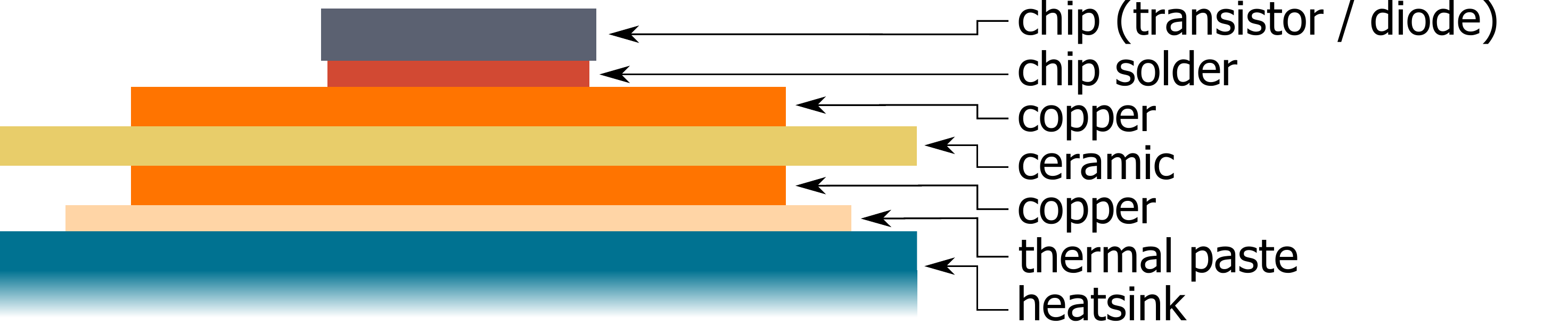}
\par\end{centering}
}
\par\end{center}%
\end{minipage}%
\begin{minipage}[t]{0.34\columnwidth}%
\begin{center}
\subfloat[Complicated layout comprising diodes and transistors\label{fig:Complicated-layout}]{\begin{centering}
\includegraphics[width=1\columnwidth]{images/powerModule_decapsulated}
\par\end{centering}
}
\par\end{center}%
\end{minipage}
\par\end{centering}
\centering{}\caption{Semikron power electronics module SK20 DGDL 065 ET}
\end{figure}

A selection of the basic discretization level is a trade-off between
growing number of compartments and a quality of the model. A layer
arrangement in one axis (Z-axis in our case) is typical for a common
power semiconductor module (Figure \ref{fig:layerArrangment}), whereas
more complicated layout comprising diodes or transistors covers the
surface (Figure \ref{fig:Complicated-layout}) in remaining two axes
(X- and Y-axis). For this reason, the suggested design of the discretization
divides into the selection of the number of compartment's layers and
selection of the fineness of the power module surface grid.

The specific discretization levels used in this paper for testing
the proposed identification method are discussed in Section \ref{sec:ValidationSyntheticData}.

\section{Compartment state space model}

The governing equation for the LTI SS model is the well known heat
transfer equation
\begin{equation}
\text{div}\left(\lambda\text{grad}T\right)+p=\rho c_{\mathrm{p}}\frac{\partial T}{\partial t},\label{eq:heatTransferEq}
\end{equation}
where $T$ stands the temperature, $p$ is the internal heat source
(with base units $[\mathrm{W/m^{3}}]$), $\text{\ensuremath{\lambda}}$
is the thermal conductivity coefficient, $\rho$ is the density of
the material, and $c_{\mathrm{p}}$ is the specific heat capacity
at constant pressure. Under the assumption of uniform constant thermal
properties of compartments, equation \eqref{eq:heatTransferEq} can
be discretized \citep{patankar2018numerical} using fully explicit
scheme into the form
\begin{equation}
T_{i,t+1}=T_{i,t}+\Delta\tau\left(\sum_{j\in\mathcal{S}_{i}}k_{i,j}(T_{j,t}-T_{i,t})+z_{i}P_{i,t}\right),\label{eq:model_discretisation}
\end{equation}
where $T_{i,t}$ and $P_{i,t}$ are the temperature and the internal
volumetric heat source inside the compartment $i$ in the discrete
time $t$, $t\in\mathbb{Z}$. The time step is denoted by $\text{\ensuremath{\Delta}}\tau$
and $\mathcal{S}_{i}$ is the set of indexes of the compartments adjacent
to the $i$th compartment such that we assume thermal coupling with
the $i$th compartment. More detail can be seen in \citep{our_iecon2019}.

The proposed compartment model \eqref{eq:model_discretisation} can
be viewed as a particular case of directed graphs. Using graph theory
\citep{bondy2007graph}, the heat transfer between compartments given
by equation \eqref{eq:model_discretisation} may be described by a
\textit{directed graph} with \textit{vertices} $T_{i}$ and \textit{directed
edges }$k_{i,j}$. Coefficients $k_{i,j}$ can be arbitrarily sorted
into a vector $\boldsymbol{k}\text{\ensuremath{\in}}\mathbb{R}^{m}$,
which corresponds to the ordering of edges in the graph. Then the
directed graph can be represented by an \textit{incidence matrix}
$\mathcal{J}$. The incidence matrix is a sparse matrix of size $n\text{\ensuremath{\times}}m$
in general, where $n$ is the number of vertices (i.e. compartments)
and $m$ is the number of edges (i.e. valid coefficients $k_{i,j}$).
The element $j_{i,l}$ of the incidence matrix $\mathcal{J}$ is defined
by the relation
\begin{equation}
j_{i,l}=\left\{ \begin{array}{ll}
1 & \textrm{\qquad if }T_{i}\textrm{ is the tail of the }l\textrm{-th edge}\\
-1 & \textrm{\qquad if }T_{i}\textrm{ is the head of the }l\textrm{-th edge}\\
0 & \textrm{\qquad otherwise}.
\end{array}\right.
\end{equation}

Introducing a temperature vector $\boldsymbol{T}_{t}=[T_{1,t},\ldots,T_{n,t}]'$,
a vector of volumetric power sources (power losses) $\boldsymbol{P}_{t}=[P_{1,t},\ldots,P_{n,t}]'$
and a parameter vector $\boldsymbol{z}=[z_{1},\ldots,z_{n}]'$, and
employing the incidence matrix $\mathcal{J}$ and the parameter vector
$\boldsymbol{k}$, discrete thermal dynamic equation \eqref{eq:model_discretisation}
can be written in the form using the parametrization by $\boldsymbol{k}$
and $\boldsymbol{z}$
\begin{eqnarray}
\boldsymbol{T}_{t+1} & = & \boldsymbol{T}_{t}-\Delta\tau\mathcal{I}\mathrm{\text{diag}(\mathcal{C}\boldsymbol{k})\mathcal{J}'}\boldsymbol{T}_{t}+\Delta\tau\mathcal{B}\mathrm{\text{diag}}(\mathcal{A}\boldsymbol{z})\boldsymbol{P}_{t}\label{eq:model_E-step_graphNetwork}\\
 & = & \boldsymbol{T}_{t}-\Delta\tau\mathcal{I}\text{diag}(\mathcal{J}'\boldsymbol{T}_{t})\mathcal{C}\boldsymbol{k}+\Delta\tau\mathcal{B}\text{diag}(\boldsymbol{P}_{t})\mathcal{A}\boldsymbol{z},\label{eq:model_M-step_graphNetwork}
\end{eqnarray}
where matrix $\text{\ensuremath{\mathcal{I}}}$ is a matrix obtained
from incidence matrix $\mathcal{J}$ by replacement $1\rightarrow0$.
Matrices and $\mathcal{A},\mathcal{B}$ and $\mathcal{C}$ in equations
\eqref{eq:model_E-step_graphNetwork} and \eqref{eq:model_M-step_graphNetwork}
are auxiliary matrices of elementary vectors. If $\text{\ensuremath{\mathcal{A}}}=I_{n}$,
$\text{\ensuremath{\mathcal{B}}}=I_{n}$, $\text{\ensuremath{\mathcal{C}}}=I_{m}$,
where $I_{n}$ is the identity matrix $n\text{\ensuremath{\times}}n$,
we get exactly the same set of equations as in \eqref{eq:model_discretisation}.

In many cases, the power losses $\boldsymbol{P}_{t}$ could be considered
only for particular compartments (e.g. for compartments corresponding
to transistors or diodes in the case of power module modeling). Furthermore
we wish to have the mesh-based compartment model described by shared
parameters. It means that selected sets of coefficients $k_{i,j}$
and $z_{i}$ are required to be identical. These presumptions are
very desirable for the following identification procedure, since the
dimension of temperature vector in the proposed mesh-based model is
relatively high.

In general, vectors $\boldsymbol{z}=[z_{1},\ldots,z_{n_{z}}]'$, $\boldsymbol{P}_{t}=[P_{1,t},\ldots,P_{n_{P},t}]'$,
and $\boldsymbol{k}=[k_{1},\ldots,k_{n_{k}}]'$ are of arbitrary lengths
$n_{z}$, $n_{P}$, and $n_{k}$ respectively. Then $\mathcal{A}$
is a matrix $n_{P}\text{\ensuremath{\times}}n_{z}$ of scaled elementary
row vectors mapping vector $\boldsymbol{z}$ to the corresponding
heat sources, $\mathcal{B}$ is a matrix $n\text{\ensuremath{\times}}n_{P}$
of scaled elementary row vectors mapping heat sources to the corresponding
compartments, and $\mathcal{C}$ is a matrix $m\text{\ensuremath{\times}}n_{k}$
of scaled elementary row vectors mapping vector $\boldsymbol{k}$
to the corresponding edges of the graph representation (i.e. mapping
vector $\boldsymbol{k}$ to the corresponding differences $T_{j,t}-T_{i,t}$).
The scale of each elementary vector corresponds to a certain weight
of transfer coefficients among compartments in the case that only
a fraction of compartment volume is occupied by the studied device,
or in the case that compartment refining is utilized during designing
of the mesh-based model. Then, these weights are dependent on the
different volumes and outer surfaces of base-size compartments and
refined compartments. In other cases the default scale is set to one.

Further, equations \eqref{eq:model_E-step_graphNetwork}\textendash \eqref{eq:model_M-step_graphNetwork}
can be rewritten in a more pleasant form. Establishing matrices $A$,
$B$ and $M$
\begin{eqnarray}
A & = & I_{n}-\Delta\tau\mathcal{I}\mathrm{\text{diag}(\mathcal{C}\boldsymbol{k})\mathcal{J}'},\label{eq:Agivenbyk}\\
B & = & \Delta\tau\mathcal{B}\mathrm{\text{diag}}(\mathcal{A}\boldsymbol{z}),\label{eq:Bgivenbyz}\\
M_{t} & = & \left[-\mathcal{I}\text{diag}(\mathcal{J}'\boldsymbol{T}_{t})\mathcal{C},\,\text{\ensuremath{\mathcal{B}}diag}(\boldsymbol{P}_{t})\mathcal{A}\right]
\end{eqnarray}
and assuming an additional zero-mean Gaussian noise $\boldsymbol{w}_{t}\text{\ensuremath{\in\ }}\mathbb{R}^{n}$
with a covariance matrix $Q$, $\boldsymbol{w}_{t}\sim\mathcal{N}\left(\boldsymbol{w}_{t}|\boldsymbol{0},Q\right)$,
equation \eqref{eq:model_E-step_graphNetwork} can be put into the
standard form of the explicit discrete LTI state equation 
\begin{eqnarray}
\boldsymbol{T}_{t+1} & = & A\boldsymbol{T}_{t}+B\boldsymbol{P}_{t}+\boldsymbol{w}_{t}\label{eq:general_E-step_graphNetwork-1}\\
 & = & \boldsymbol{T}_{t}+\Delta\tau M_{t}\boldsymbol{\theta}+\boldsymbol{w}_{t},\label{eq:general_M-step_graphNetwork-1}\\
\text{\ensuremath{\hphantom{\boldsymbol{\boldsymbol{}_{t+1}}}}}\boldsymbol{\theta} & = & \left[\boldsymbol{k}',\boldsymbol{z}'\right]'.
\end{eqnarray}
with the state vector $\boldsymbol{T}_{\bullet}$ and the input vector
$\boldsymbol{P}_{\bullet}$. Equation \eqref{eq:general_M-step_graphNetwork-1}
is a notation enabling to use a least squares method for estimating
unknown parameter vectors $\boldsymbol{k}$ and $\boldsymbol{z}$
of the proposed model.

For completeness of the model, we define the vector of measured temperatures
as $\boldsymbol{y}_{t}$ and observation model (output equation)
\begin{equation}
\boldsymbol{y}_{t}=C\boldsymbol{T}_{t}+\boldsymbol{v}_{t},\label{eq:SS-outputEquation}
\end{equation}
where $C\in\mathbb{R}^{n_{y}\times n}$ is the matrix comprising elementary
row vectors corresponding to the indices of $n_{y}$ measured (observed)
compartment temperatures, and $\boldsymbol{v}_{t}$ is a zero-mean
Gaussian noise with a covariance matrix $R$, $\boldsymbol{v}_{t}\sim\mathcal{N}\left(\boldsymbol{v}_{t}|\boldsymbol{0};R\right).$
The length of the measurement vector is denoted by $N,$ $t\in1:N.$
Then, equations \eqref{eq:general_E-step_graphNetwork-1} and \eqref{eq:SS-outputEquation}
form a standard discrete LTI SS model with unknown parameters $\boldsymbol{\theta}$.

\section{Expectation-maximization algorithm}

The Expectation-Maximization (EM) algorithm \citep{dempster1977maximum}
is a standard technique that allows to estimate model parameters from
data sets with missing or hidden variables. Its application for the
identification of the proposed model \eqref{eq:general_E-step_graphNetwork-1}\textendash \eqref{eq:SS-outputEquation}
is now reviewed.

The objective of EM algorithm is to maximize the log likelihood of
the measured data
\begin{equation}
\log p\left(Y|P,\boldsymbol{\theta}\right)=\log\int\limits _{T}p\left(T,Y|P,\boldsymbol{\theta}\right)dT,\label{eq:EMalg-logMarginalLikelihood}
\end{equation}
where $T=\left\{ \boldsymbol{T}_{1},\cdots,\boldsymbol{T}_{N}\right\} $,
$P=\left\{ \boldsymbol{P}_{1},\cdots,\boldsymbol{P}_{N}\right\} $,
$Y=\left\{ \boldsymbol{y}_{1},\cdots,\boldsymbol{y}_{N}\right\} $.
In essence, the algorithm is proposed to approximate the correct marginal
likelihood approach by iterative maximization of its lower bound.
The lower bound $\mathcal{F}\left(\mathcal{Q},\boldsymbol{\theta}\right)$
is derived \citep{roweis2000algorithm}  using any distribution $\Phi\left(T\right)$
as
\begin{eqnarray}
\log\int\limits _{T}p\left(T,Y|P,\boldsymbol{\theta}\right)dT=\log\int\limits _{T}\Phi\left(T\right)\frac{p\left(T,Y|P,\boldsymbol{\theta}\right)}{\Phi\left(T\right)}dT=\hspace{2cm}\nonumber \\
\hspace{1cm}=\log E_{\Phi\left(T\right)}\left[\frac{p\left(T,Y|P,\boldsymbol{\theta}\right)}{\Phi\left(T\right)}\right]\geq E_{\Phi\left(T\right)}\left[\log\frac{p\left(T,Y|P,\boldsymbol{\theta}\right)}{\Phi\left(T\right)}\right]=\hspace{1cm}\\
\vphantom{\int\limits _{X}}\hspace{2cm}=E_{\Phi\left(T\right)}\left[\log p\left(T,Y|P,\boldsymbol{\theta}\right)\right]-E_{\Phi\left(T\right)}\left[\log\Phi\left(T\right)\right]=\mathcal{F}\left(\Phi,\boldsymbol{\theta}\right).\nonumber 
\end{eqnarray}
To find the maximum likelihood (ML) estimate of unknown parameters
$\boldsymbol{\theta}$, the EM algorithm seeks to maximize the lower
bound $\mathcal{F}\left(\Phi,\boldsymbol{\theta}\right)$ of the observed
data marginal likelihood \eqref{eq:EMalg-logMarginalLikelihood} by
alternating so called Expectation step (E-step) and Maximization step
(M-step). With the aim to identify the proposed LTI SS model, we discuss
these steps in the following text in more detail.\textit{}

\subsection{Expectation step \textendash{} Rauch Tung Striebel Smoother}

In the $r$-th E-step, the lower bound $\mathcal{F}(\Phi,\boldsymbol{\theta}^{r-1})$
is maximized with respect to the distribution $\Phi$ holding fixed
parameter vector $\boldsymbol{\theta}^{r-1}$. Assuming that we have
some parameter values $\boldsymbol{\theta}^{r-1}$ available from
the previous M-step, it is possible to show, that the desired distribution
$\Phi^{r}\left(T\right)$ is exactly the conditional distribution
of $T$ \citep{roweis2000algorithm}
\begin{equation}
\Phi^{r}\left(T\right)=p\left(T|Y,P,\boldsymbol{\theta}^{r-1}\right)\label{eq:conditionalDistributionT}
\end{equation}

Since for the known values of the parameter vector $\boldsymbol{\theta}^{r-1}$,
i.e. for known values of matrix $A$ and $B$ given by \eqref{eq:Agivenbyk}
and \eqref{eq:Bgivenbyz}, the system \eqref{eq:general_E-step_graphNetwork-1},\eqref{eq:SS-outputEquation}
forms a state space model, the full distribution of all temperatures
\eqref{eq:conditionalDistributionT} can be determined by Rauch Tung
Striebel Smoother (RTSS) \citep{rauch1965maximum}. The RTSS is a
two-pass algorithm (Algorithm \ref{alg:original-RTS}) for fixed interval
smoothing, where the first pass is the regular forward Kalman filter
and the second pass is the backward smoother.

The output of the RTSS is the smoothed posterior Gaussian distribution
$p(\boldsymbol{T}_{t}|P,\boldsymbol{\theta}^{r-1},Q^{r-1},R)$. Moreover
theRTSS purveys the smoothed posterior joint distribution 
\begin{equation}
p\left(\left[\hspace{-0.15cm}\begin{array}{c}
\boldsymbol{T}_{t+1}\\
\boldsymbol{T}_{t}
\end{array}\hspace{-0.15cm}\right]|P,\boldsymbol{\theta}^{r-1},Q^{r-1},R\right)=\mathcal{N}\left(\left[\hspace{-0.15cm}\begin{array}{c}
\boldsymbol{T}_{t+1}\\
\boldsymbol{T}_{t}
\end{array}\hspace{-0.15cm}\right]|\left[\hspace{-0.15cm}\begin{array}{c}
\boldsymbol{x}_{t+1}^{N}\\
\boldsymbol{x}_{t}^{N}
\end{array}\hspace{-0.15cm}\right],\left[\hspace{-0.15cm}\begin{array}{cc}
V_{t+1}^{N} & V_{t+1,t}^{N}\\
\left(V_{t+1,t}^{N}\right)' & V_{t}^{N}
\end{array}\hspace{-0.15cm}\right]\right),\label{eq:posteriorSmoothedDistribution}
\end{equation}
where the notation from the Algorithm \ref{alg:original-RTS} is used.
Note that the temperature random variable is marked by $\boldsymbol{T}_{\bullet}$,
whereas the smoothed estimate (or the expected value in other words)
by $\boldsymbol{x}_{\bullet}^{N}$.

As can be seen, the posterior distributions determined by RTSS are
also conditioned by covariance matrices $Q^{r-1}$ and $R$. These
matrices can be either known in many cases and fixed by user or they
can be added into the identification process. In this paper, we assume
covariance matrix $R$ of measurement noise to be known and we incorporate
the identification of process noise covariance matrix $Q$ into the
estimation procedure. Therefore, $Q^{r-1}$ is available in the $r$-th
E-step similarly to parameters vector $\boldsymbol{\theta}^{r-1}$.
The estimation of $Q^{r-1}$ is the objective of the previous maximization
step discussed in subsection \ref{subsec:Maximization-step}.

\subsection{Speeding up of E-step using steady-state covariances}

The time and memory burdens of RTSS directly depend on the dimension
of state space vector $\boldsymbol{T}_{t}$ (we assume the dimension
of vector $\boldsymbol{T}_{t}$ much larger than dimensions of vectors
$\boldsymbol{P}_{t}$ and $\boldsymbol{y}_{t}$) and on the number
of measurements $N$. Since we are using the model with hundreds or
even thousands of compartments (corresponding to the dimension of
$\boldsymbol{T}_{t}$ and each dimension of covariance matrices $V_{t+1}^{\text{\ensuremath{\bullet}}}$),
especially the inversion $(V_{t+1}^{t})^{-1}$(line \ref{line:inversionCovarianceRST}
in Algorithm \ref{alg:original-RTS}) in each for-cycle iteration
of the backward pass is very time consuming apart from a large matrix
multiplication in the remaining parts of RTSS. Besides, storing covariance
matrices $V_{t}^{t}$ and $V_{t+1}^{t}$ in the forward pass (necessary
for backward pass) is strongly memory-consuming as the number of observation
$N$ increases.\\
\begin{minipage}[t]{0.42\columnwidth}%
\begin{algorithm}[H]
\begin{algorithmic}[1]

\STATE  \textbf{input} $A,$~$B,$~$C,$~$Q,$~$R,$

\hspace{1cm}$\boldsymbol{x}_{1}^{1}=\boldsymbol{T}_{1},$~$\boldsymbol{P}_{t},$~$\boldsymbol{y}_{t}$,~$N$

~\vphantom{$V_{S}^{-}$:}

~\vphantom{$\left(CV_{S}^{-}C'+R\right)^{-1}$}

\STATE  \textbf{for} $t=1:1:N-1$

\STATE  \quad{}$\boldsymbol{x}_{t+1}^{t}=A\mathbf{x}_{t}^{t}+B\boldsymbol{P}_{t}$

\STATE  \quad{}$V_{t+1}^{t}=AV_{t}^{t}A'+Q$

\STATE  \quad{}$K_{t+1}=V_{t+1}^{t}C'(CV_{t+1}^{t}C'+R)^{-1}$

\STATE  \quad{}$V_{t+1}^{t+1}=V_{t+1}^{t}-K_{t+1}CV_{t+1}^{t}$

\STATE  \quad{}$\boldsymbol{x}_{t+1}^{t+1}=\boldsymbol{x}_{t+1}^{t}+K_{t+1}(\boldsymbol{y}_{t+1}-C\boldsymbol{x}_{t+1}^{t})$

\STATE  \textbf{end for}

~\vphantom{$V_{S}^{N}:$}

\STATE  \textbf{for} $t=N-1:-1:1$

\STATE  \quad{}$J_{t}=V_{t}^{t}A'(V_{t+1}^{t})^{-1}$\label{line:inversionCovarianceRST}

\STATE  \quad{}$V_{t}^{N}=V_{t}^{t}+J_{t}(V_{t+1}^{N}-V_{t+1}^{t})J_{t}'$

\STATE  \quad{}$V_{t+1,t}^{N}=V_{t+1}^{N}J_{t}'$

\STATE  \quad{}$\boldsymbol{x}_{t}^{N}=\boldsymbol{x}_{t}^{t}+J_{t}(\boldsymbol{x}_{t+1}^{N}-A\boldsymbol{x}_{t}^{t}-B\boldsymbol{P}_{t})$

\STATE  \textbf{end for}

\end{algorithmic}

\caption{original RTSS\label{alg:original-RTS}}
\end{algorithm}
\end{minipage}\hspace{0.018\columnwidth}%
\begin{minipage}[t]{0.56\columnwidth}%
\begin{algorithm}[H]
\begin{algorithmic}[1]

\STATE  \textbf{input} $A,$~$B,$~$C,$~$Q,$~$R,$

\hspace{1cm}$\boldsymbol{x}_{1}^{1}=\boldsymbol{T}_{1},$~$\boldsymbol{P}_{t},$~$\boldsymbol{y}_{t}$,~$N$

\vspace{0.36cm}

\begin{raggedright}
\STATE  compute Riccati equation for $V_{S}^{-}$:\label{line:steady-Riccati}
\par\end{raggedright}
\begin{centering}
$V_{\ensuremath{S}}^{-}=AV_{S}A'-AV_{S}^{-}C'\left(CV_{S}^{-}C'+R\right)^{-1}CV_{S}^{-}A'+Q$
\par\end{centering}
\STATE  $K_{S}=V_{S}^{-}C'\left(CV_{S}^{-}C'+R\right)^{-1}$

\STATE  $V_{S}^{+}=\left(I_{n}-K_{S}C\right)V_{S}^{-}$

\STATE  \textbf{for} $t=1:1:N-1$

\STATE  \quad{}$\boldsymbol{x}_{t+1}^{t}=A\mathbf{x}_{t}^{t}+B\boldsymbol{P}_{t}$

\STATE  \quad{}$\boldsymbol{x}_{t+1}^{t+1}=\boldsymbol{x}_{t+1}^{t}+K_{S}(\boldsymbol{y}_{t+1}-C\boldsymbol{x}_{t+1}^{t})$

\STATE  \textbf{end for}

~

\STATE $J_{S}=V_{S}^{+}A'(V_{S}^{-})^{-1}$
\begin{raggedright}
\STATE  compute Lyapunov equation for $V_{S}^{N}:$\label{line:steady-lyapunov}
\par\end{raggedright}
\begin{centering}
$V_{S}^{N}=J_{S}V_{S}^{N}J_{S}'+\left(V_{S}^{+}-J_{S}V_{S}^{-}J_{S}'\right)$
\par\end{centering}
\STATE  \textbf{for} $t=N-1:-1:1$

\STATE  \quad{}$\boldsymbol{x}_{t}^{N}=\boldsymbol{x}_{t}^{t}+J_{S}(\boldsymbol{x}_{t+1}^{N}-A\boldsymbol{x}_{t}^{t}-B\boldsymbol{P}_{t})$

\STATE  \textbf{end for}

\end{algorithmic}

\caption{RTSS with steady covariances\label{alg:RTSS-with-steady}}
\end{algorithm}
\end{minipage}

\vspace{0.1cm}

For these reasons, we suggest to use steady covariance matrices in
the RTSS which significantly reduces the computational requirements
\citep{crassidis2011optimal}. The implementation of the RTSS with
steady covariances is shown in Algorithm \ref{alg:RTSS-with-steady}.

Riccati equation on line \ref{line:steady-Riccati} and Lyapunov equation
on line \ref{line:steady-lyapunov} of Algorithm \ref{alg:RTSS-with-steady}
can be evaluated by direct method or solvers (e.g. \texttt{idare}
and \texttt{dlyap} Matlab's in-build function) or iteratively using
e.g. Newton techniques. Specifically in our case, we employed Modified
Newton method for discrete-time algebraic Riccati equations \citep{sima2014riccati}
and Matlab's function \texttt{dlyap} for solving discrete-time Lyapunov
equations.

For an effective implementation, it is sufficient to collect only
few statistics of relatively small size (square of the number of compartments)
for the following M-step. The necessary statistics utilized in the
M-step and obtained from RTSS using steady covariance matrices are
\begin{eqnarray}
XX'\equiv(N-1)V_{S}^{N}+\sum_{t=1}^{N-1}\boldsymbol{x}_{t}^{N}\left(\boldsymbol{x}_{t}^{N}\right)' &  & XU'\equiv\sum_{t=1}^{N-1}\boldsymbol{x}_{t}^{N}\boldsymbol{P}_{t}'\nonumber \\
ZZ'\equiv(N-1)V_{S}^{N}+\sum_{t=1}^{N-1}\boldsymbol{x}_{t+1}^{N}\left(\boldsymbol{x}_{t+1}^{N}\right)' &  & ZU'\equiv\sum_{t=1}^{N-1}\boldsymbol{x}_{t+1}^{N}\boldsymbol{P}_{t}'\label{eq:RTSSstatistics}\\
XZ'\equiv(N-1)J_{S}\left(V_{S}^{N}\right)'+\sum_{t=1}^{N-1}\boldsymbol{x}_{t}^{N}\left(\boldsymbol{x}_{t+1}^{N}\right)' & \hspace{1cm} & UU'\equiv\sum_{t=1}^{N-1}\boldsymbol{P}_{t}\boldsymbol{P}_{t}'\nonumber 
\end{eqnarray}

For comparison, the form of statistics derived by the original full
RTSS (Algorithm \ref{alg:original-RTS}) and utilizable for the M-step
can be found in \citep{our_iecon2019}.

\subsection{Maximization step \textendash{} Maximum Likelihood Estimate \label{subsec:Maximization-step}}

In the $r$-th M-step, the lower bound $\mathcal{F}(\Phi^{r},\boldsymbol{\theta})$
is maximized with respect to the unknown model parameters $\boldsymbol{\theta}$
and noise covariance matrix $Q$ holding distribution $\Phi^{r}(T)$
fixed. The distribution $\Phi^{r}(T)$ is the smoothed posterior distribution
\eqref{eq:posteriorSmoothedDistribution} evaluated in the previous
E-step of EM algorithm. Then, the new updates of parameters $\boldsymbol{\theta}^{r}$
and noise covariance matrix $Q^{r}$are given by 
\begin{eqnarray}
\boldsymbol{\theta}^{r},Q^{r}\text{=} & \arg\max_{\boldsymbol{\theta},Q}\sum\limits _{1}^{N-1}E_{\Phi^{r}(T)} & \Bigl\{\ln|Q^{-1}|\ensuremath{|R^{-1}|}-\left(\Delta\boldsymbol{T}_{t}-M_{t}\boldsymbol{\theta}\right)'Q^{-1}\left(\Delta\boldsymbol{T}_{t}-M_{t}\boldsymbol{\theta}\right)+\nonumber \\
 &  & \vphantom{\ln|Q^{-1}|\ensuremath{|R^{-1}|}-\left(\Delta\boldsymbol{T}_{t}-M_{t}\boldsymbol{\theta}\right)'Q^{-1}\left(\Delta\boldsymbol{T}_{t}-M_{t}\boldsymbol{\theta}\right)}-\text{ \ensuremath{\left(\boldsymbol{y}_{t}-C\boldsymbol{T}_{t}\right)'}}R^{-1}\text{ \ensuremath{\left(\boldsymbol{y}_{t}-C\boldsymbol{T}_{t}\right)}}\Bigr\}\label{eq:Mstep_argMaxFull}
\end{eqnarray}
where $\Delta\boldsymbol{T}_{t}=\Delta\tau^{-1}(\boldsymbol{T}_{t+1}-\boldsymbol{T}_{t})$,
$E_{\Phi^{r}(T)}(\cdot)$ stands for the expectation value with respect
to the distribution \eqref{eq:posteriorSmoothedDistribution} and
$|\cdot|$ denotes the determinant of the particular matrix. 

It is easy to show, that the ML estimator of \eqref{eq:Mstep_argMaxFull}
is of the form
\begin{eqnarray}
\boldsymbol{\theta}^{r}=\left(\sum_{t=1}^{N-1}E_{\Phi^{r}(T)}\left\{ M_{t}'\left(Q^{r}\right)^{-1}M_{t}\right\} \right)^{-1}\sum_{t=1}^{N-1}E_{\Phi^{r}(T)}\left\{ M_{t}'\left(Q^{r}\right)^{-1}\Delta\boldsymbol{T}_{t}\right\} \label{eq:MLE_theta}\\
Q_{\textrm{full}}^{r}\text{=}\frac{1}{N-1}\hspace{-0.15cm}\sum\limits _{t=1}^{N-1}\hspace{-0.1cm}E_{\Phi^{r}(T)}\Bigl\{\hspace{-0.05cm}\Delta\boldsymbol{T}_{t}\Delta\boldsymbol{T}_{t}'\hspace{-0.05cm}-\hspace{-0.05cm}\Delta\boldsymbol{T}_{t}\hspace{-0.05cm}\left(M_{t}\boldsymbol{\theta}^{r}\right)'\hspace{-0.12cm}-\hspace{-0.05cm}M_{t}\boldsymbol{\theta}^{r}\hspace{-0.05cm}\left(\Delta\boldsymbol{T}_{t}\right)'\hspace{-0.12cm}+\hspace{-0.05cm}M_{t}\boldsymbol{\theta}^{r}\hspace{-0.05cm}\left(M_{t}\boldsymbol{\theta}^{r}\right)'\hspace{-0.05cm}\Bigr\}\label{eq:MLE_Q}
\end{eqnarray}
where the individual expected terms can be expressed using only statistics
\eqref{eq:RTSSstatistics} obtained by RTSS in the previous $r$-th
E-step. These terms are given in \ref{sec:Appendix_expected-terms-for-Mstep}
in the detail.

The computational problem lies in the mutual cross dependency of $\boldsymbol{\theta}^{r}$
on $Q^{r}$ and vice versa. This obstacle is connected with the structure
of desired covariance matrix $Q$ and may vanish in some particular
case. Moreover, the number of elements in the covariance matrix $Q$
is much higher than dimension of parameters vector $\boldsymbol{\theta}$
and thus a regularization of the problem (a shrinkage of covariance
matrix estimation) is greatly desirable. Therefore, we investigate
carefully the structure of the covariance matrix $Q$ now and discuss
possible solutions.

\subsection{Structure of process noise covariance matrix \label{subsec:Structure-of-processCovariance}}
\begin{enumerate}
\item An easy approach is to assume the covariance matrix $Q$ in the diagonal
form with a constant on the diagonal, $Q\overset{!}{=}qI_{n}$. In
such case, equation \eqref{eq:MLE_theta} can be simplified, since
the term $\left(qI_{n}\right)^{-1}$ is possible to completely eliminate
from the expression for $\boldsymbol{\theta}^{r}$ \eqref{eq:MLE_theta}.
Thus the evaluation of $\boldsymbol{\theta}^{r}$ is not dependent
on the constrained covariance matrix $Q$ and can be directly executed.
The formulation of M-step is then similar to ordinary least squares
method but with proper considering of expected values. The formula
for computation of the covariance matrix $q^{r}I_{n}$, or scalar
value $q^{r}$ actually, reads
\begin{equation}
q^{r}=\frac{1}{n}\Tr\left(Q_{\textrm{full}}^{r}\right).
\end{equation}
This case together with specific form of equations \eqref{eq:MLE_theta}
and \eqref{eq:MLE_Q} is described by authors in \citep{our_iecon2019}.
\item There exist several other constraints on the covariance matrix $Q$,
where direct derivation of the estimator is feasible. One representative
of this group is non-homogeneous diagonal covariance matrix (compare
with weighted least squares method), $Q\overset{!}{=}\text{diag}\left(\boldsymbol{q}\right)$,
where $\boldsymbol{q}=[q_{1},\ldots,q_{n}]'$. In such case, the form
of ML estimators of $\boldsymbol{\theta}^{r}$ and $Q_{\textrm{diag}}^{r}$
remain as in \eqref{eq:MLE_theta} and \eqref{eq:MLE_Q}, only with
consideration that non-diagonal elements of $Q^{r}$ are zeros,
\begin{equation}
\boldsymbol{q}^{r}\text{=}\text{diag}\left(Q_{\textrm{full}}^{r}\right).
\end{equation}
The mutual cross dependency of $\boldsymbol{\theta}^{r}$ and $Q_{\textrm{diag}}^{r}$can
be overcome by employing the previous estimation of the covariance
matrix $Q_{\textrm{diag}}^{r-1}$ in \eqref{eq:MLE_theta} (i.e. utilizing
ML estimator of $Q$ from the $r-1$-th M-step of the EM algorithm
for new update of $\boldsymbol{\theta}^{r}$). Thereafter, the new
update of $\boldsymbol{\theta}^{r}$ can be used for evaluation of
\eqref{eq:MLE_Q}.
\item A constraint on the covariance matrix $Q$ can be formulated in the
``infeasible'' way, where the estimator $Q_{\alpha LL'+\beta I}^{r}$
cannot be expressed analytically in the explicit form. The proper
approach is to design an optimization task using e.g. method of Lagrange
multipliers as the way how to cope with constraints on the covariance
matrix \citep{mader2014numerically}. The other approach is to use
an approximate solution. We suggest to declare the nearest (in the
sense of Frobenius norm) constrained covariance matrix to the full
ML estimator \eqref{eq:MLE_Q} as the approximate constrained estimator
\begin{equation}
Q_{\alpha LL'+\beta I}^{r}=\arg\min_{Q_{\alpha LL'+\beta I}}\left\{ \left\Vert Q_{\textrm{full}}^{r}-Q_{\alpha LL'+\beta I}\right\Vert _{\mathcal{F}}\right\} .\label{eq:constraintCov_argimin}
\end{equation}
The specific ``infeasible'' constraint on the covariance matrix,
which we investigate in this contribution, is of the form
\begin{equation}
Q_{\alpha LL'+\beta I}^{r}=\alpha^{r}LL'+\beta^{r}I_{n},
\end{equation}
where $\alpha^{r},\beta^{r}>0$ are estimated optimal parameters and
matrix $L\in\mathbb{R}^{n\times n}$ is a predefined fixed constant
matrix. The approximate solution \eqref{eq:constraintCov_argimin}
is then easy to write using least squares method as
\begin{equation}
\left[\begin{array}{c}
\alpha^{r}\\
\beta^{r}
\end{array}\right]=\left(F_{Q}'F_{Q}\right)^{-1}F_{Q}'\text{vec}\left(Q_{\textrm{full}}^{r}\right),
\end{equation}
where matrix $F_{Q}=\left[\begin{array}{cc}
\text{vec}(LL') & \text{vec}(I_{n})\end{array}\right]$ and $\text{vec}(\cdot)$ is operator of vectorization stacking the
columns of the matrix on top of one another.
\end{enumerate}
These three various structures of the covariance matrix, the convergence
properties and their influence on the quality of results are analyzed
in Section \ref{sec:ComparisonCovarianceStructure} in the detail.

\section{Validation of the proposed method on synthetic data \label{sec:ValidationSyntheticData}}

\begin{figure}
\begin{centering}
\begin{minipage}[t]{0.47\textwidth}%
\begin{center}
\subfloat[3D visualization of the model\label{fig:layerArrangment-1}]{\begin{raggedright}
\vphantom{\includegraphics[width=1.042\textwidth]{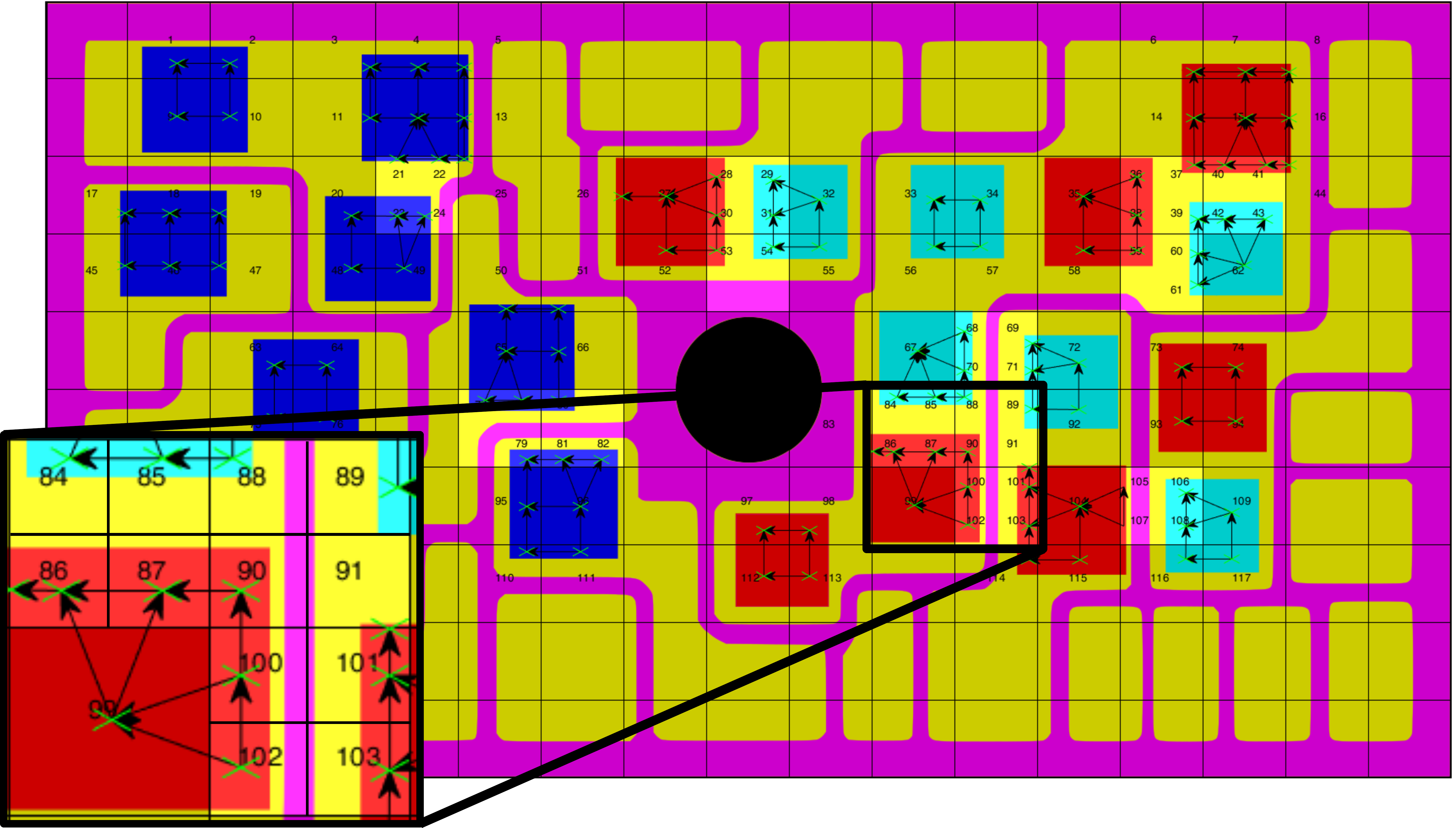}}\includegraphics[width=0.96\columnwidth]{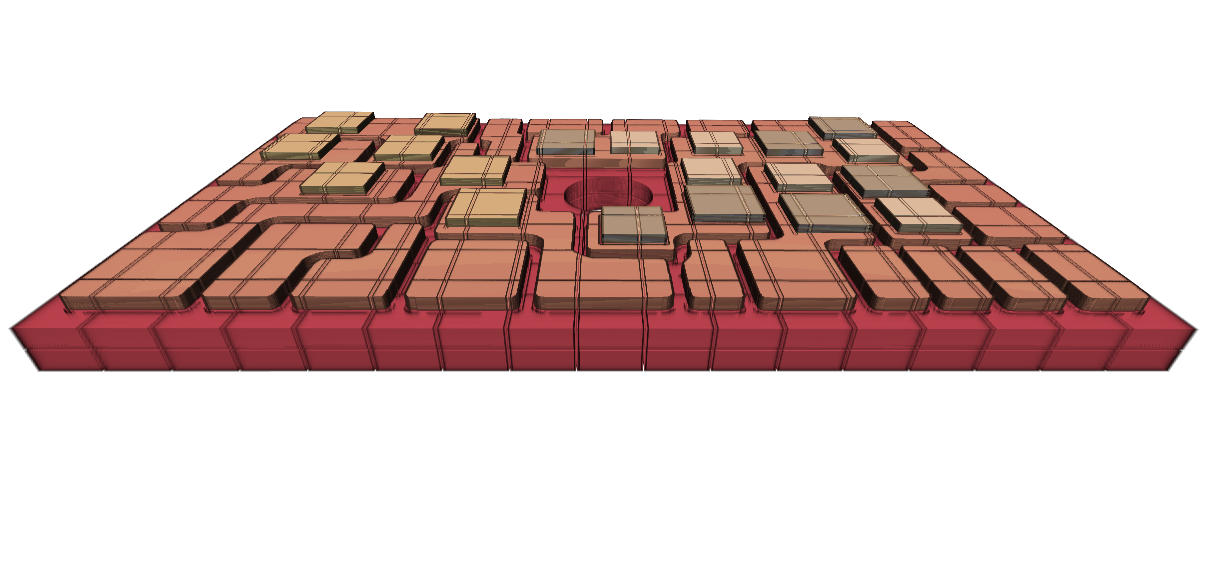}
\par\end{raggedright}
}
\par\end{center}%
\end{minipage}%
\begin{minipage}[t]{0.49\textwidth}%
\begin{center}
\subfloat[First compartments' layer of the model (refined area is zoomed)\label{fig:Complicated-layout-1}]{\begin{raggedleft}
\includegraphics[width=0.97\columnwidth]{images/1layer}
\par\end{raggedleft}
}
\par\end{center}%
\end{minipage}
\par\end{centering}
\centering{}\caption{Proposed mesh-based compartment model}
\end{figure}

The performance of the proposed identification method of mesh-based
compartment models is tested on generated data. For demonstration
purpose, we use a mesh-based compartment model inspired by physical
properties of real IGBT (insulated-gate bipolar transistor) three
phase power module SK20 DGDL 065 ET (Figure \ref{fig:Complicated-layout}).
The discretization level of the toy model is selected as follows:
four layers of compartments are used in Z-axis and the basic grid
of $17\text{\ensuremath{\times}}10$ compartments is used for each
layer in X-Y plane. Using this level of discretization, the size of
basic compartments corresponds to covering surface of size cca $3$
mm $\times3$ mm of the real power module. The total number of compartments
of the model is 817, specifically the first upper layer contains 117
compartments (caused by neglecting of surface, where no transistor,
diode or rectifier exists, and on the contrary, refining some critical
areas of the first layer \textendash{} Figure \ref{fig:Complicated-layout-1}),
the second layer contains 359 compartments (caused by refining), the
third and the fourth 170 (only basic grid used) and the last remaining
compartment is employed for the ambient temperature modeling. The
dynamics of the last compartment (representing the ambient temperature)
is dependent only on the previous value of the ambient temperature.

\subsection{Convergence of parameters \label{subsec:Convergence-of-parameters}}

The convergence of parameters is studied in this subsection. We test
two possibilities of parametrization of state matrix $A$. In the
first case, we employ 12 parameters, which are specified in Table
\ref{tab:Parametrization} and call this parametrization as weakly
shared. The second studied model employs parametrization using only
5 parameters for description of matrix $A$, which we call strongly
shared parametrization. Further there is one parameter connected with
power losses $P_{i,t}$ in compartments representing IGBTs of the
real module ($\boldsymbol{z}\in\mathbb{R}^{1}$) and parameters describing
the estimator of state noise covariance matrix $Q$.

\begin{table}[h]
\begin{centering}
\begin{tabular}{|c|c|>{\centering}p{3cm}|c|}
\hline 
\textbf{weakly shared parameters} & \textbf{true value} & \textbf{strongly shared parameters} & \textbf{true value}\tabularnewline
\hline 
IGBT (layer 1) $\rightleftarrows$ IGBT (layer 1) & 0.035 & \multirow{3}{3cm}{\centering{}$k_{1}$} & \multirow{3}{*}{0.025}\tabularnewline
\cline{1-2} \cline{2-2} 
diode (layer 1) $\rightleftarrows$ diode (layer 1) & 0.015 &  & \tabularnewline
\cline{1-2} \cline{2-2} 
rectifier (layer 1) $\rightleftarrows$ rectifier (layer 1) & 0.024 &  & \tabularnewline
\hline 
layer 2 (Cu layer) $\rightleftarrows$ layer 2 (Cu layer) & 0.022 & \multirow{3}{3cm}{\centering{}$k_{2}$} & \multirow{3}{*}{0.029}\tabularnewline
\cline{1-2} \cline{2-2} 
layer 3 $\rightleftarrows$ layer 3 & 0.044 &  & \tabularnewline
\cline{1-2} \cline{2-2} 
layer 4 $\rightleftarrows$ layer 4 & 0.020 &  & \tabularnewline
\hline 
IGBT (layer 1) $\rightleftarrows$ layer 2 (Cu layer) & 0.056 & \multirow{3}{3cm}{\centering{}$k_{3}$} & \multirow{3}{*}{0.053}\tabularnewline
\cline{1-2} \cline{2-2} 
diode (layer 1) $\rightleftarrows$ layer 2 (Cu layer) & 0.052 &  & \tabularnewline
\cline{1-2} \cline{2-2} 
rectifier (layer 1) $\rightleftarrows$ layer 2 (Cu layer) & 0.052 &  & \tabularnewline
\hline 
layer 2 (Cu layer) $\rightleftarrows$ layer 3 & 0.047 & \multirow{2}{3cm}{\centering{}$k_{4}$} & \multirow{2}{*}{0.055}\tabularnewline
\cline{1-2} \cline{2-2} 
layer 3 $\rightleftarrows$ layer 4 & 0.062 &  & \tabularnewline
\hline 
layer 4 $\rightleftarrows$ ambient temperature & 0.020 & \multirow{1}{3cm}{\centering{}$k_{5}$} & 0.020\tabularnewline
\hline 
\end{tabular}
\par\end{centering}
\caption{Parameters of studied synthetic models\label{tab:Parametrization}}

\end{table}

For identification purposes, the values of temperatures in compartments
corresponding to selected IGBTs in the real power module (specifically
40 compartments out of all 117 compartments in the first layer), temperature
of one selected compartment in layer 4 representing temperature sensor
in the real power module and the ambient temperature represented by
last compartment are observed according observation model \eqref{eq:SS-outputEquation}.
The input vector $\boldsymbol{P}_{1:N}$ is also assumed to be known.
In this subsection, we know exactly the structure of both toy models
(using weakly and strongly shared parametrization) and the true values
of all parameters which we want to identified. Moreover, the process
noise is neglected for better comparison.

\begin{figure}
\begin{centering}
\begin{minipage}[t]{0.49\columnwidth}%
\begin{center}
\subfloat[Weakly shared parameters]{\includegraphics[width=1\textwidth]{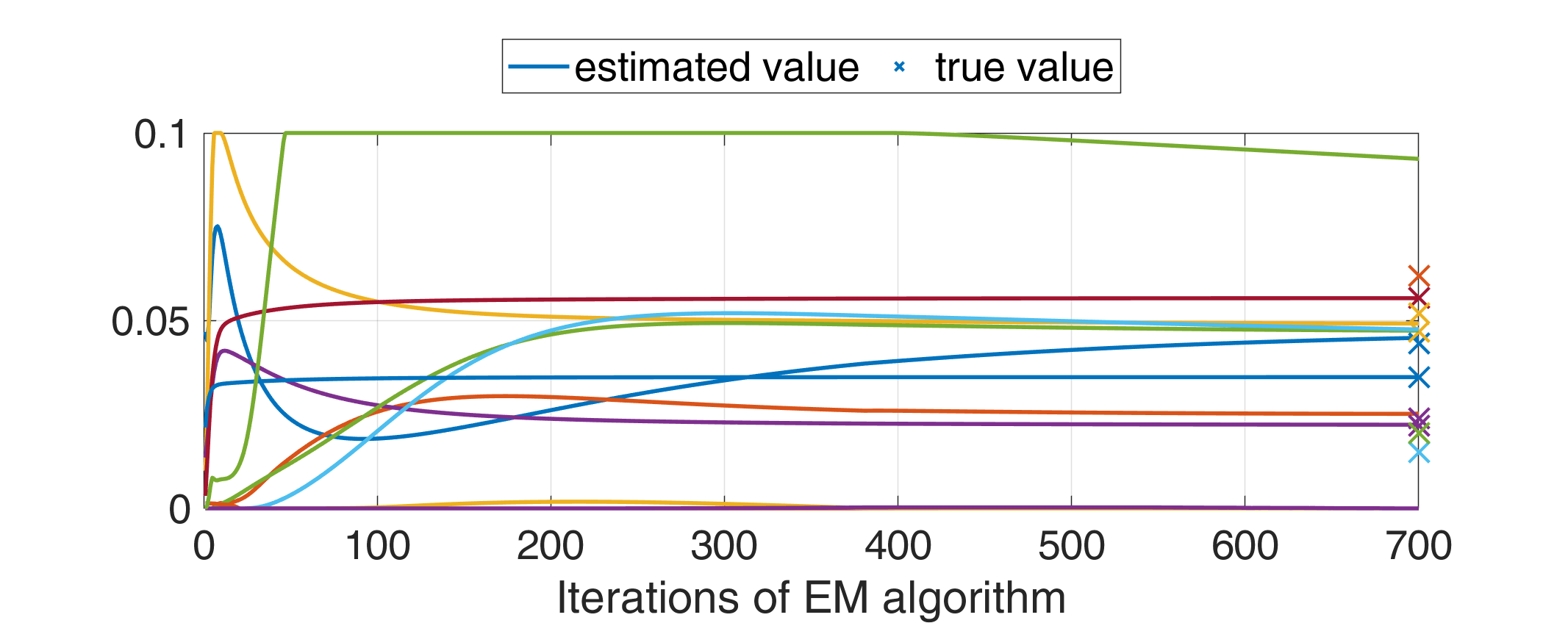}

}
\par\end{center}%
\end{minipage}%
\begin{minipage}[t]{0.49\columnwidth}%
\begin{center}
\subfloat[Strongly shared parameters]{\includegraphics[width=1\textwidth]{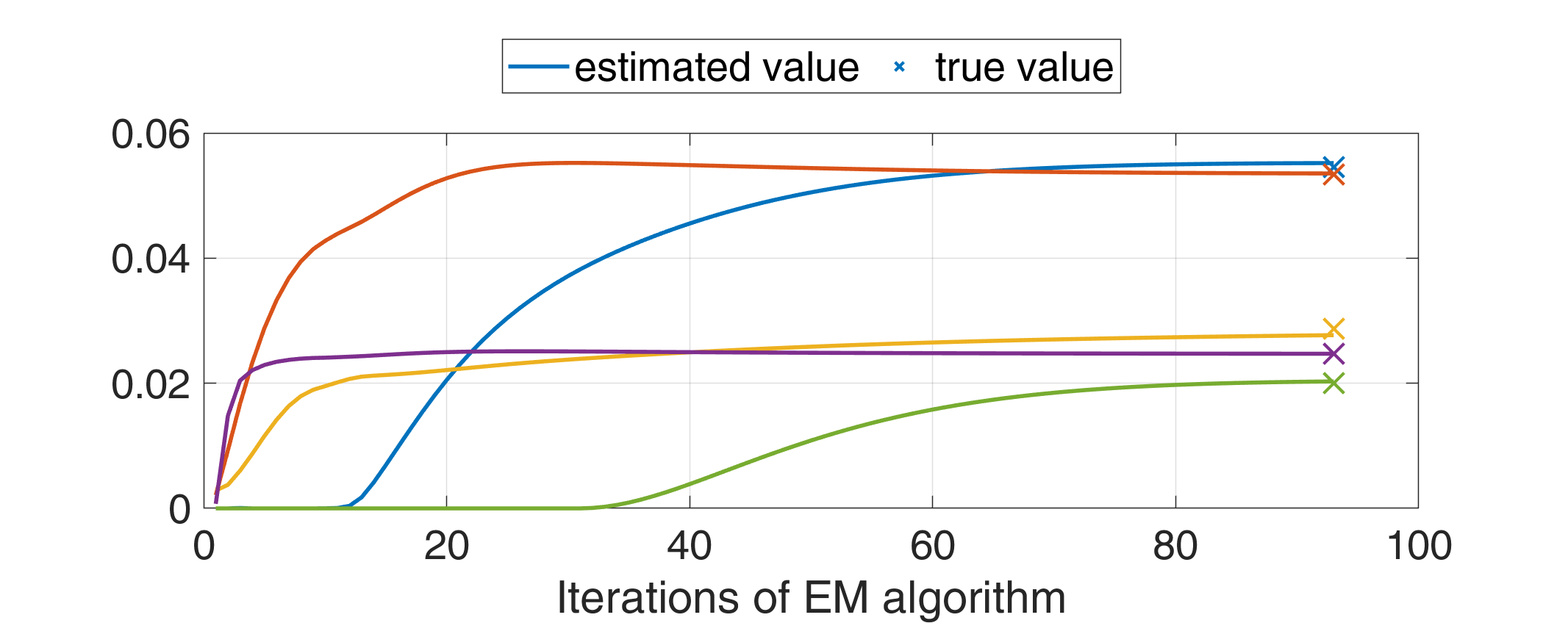}

}
\par\end{center}%
\end{minipage}
\par\end{centering}
\centering{}\caption{Convergence of values of parameters vector $\boldsymbol{k}$\label{fig:Convergence-of-parameters}}
\end{figure}

\begin{figure}[H]
\begin{centering}
\includegraphics[width=0.5\textwidth]{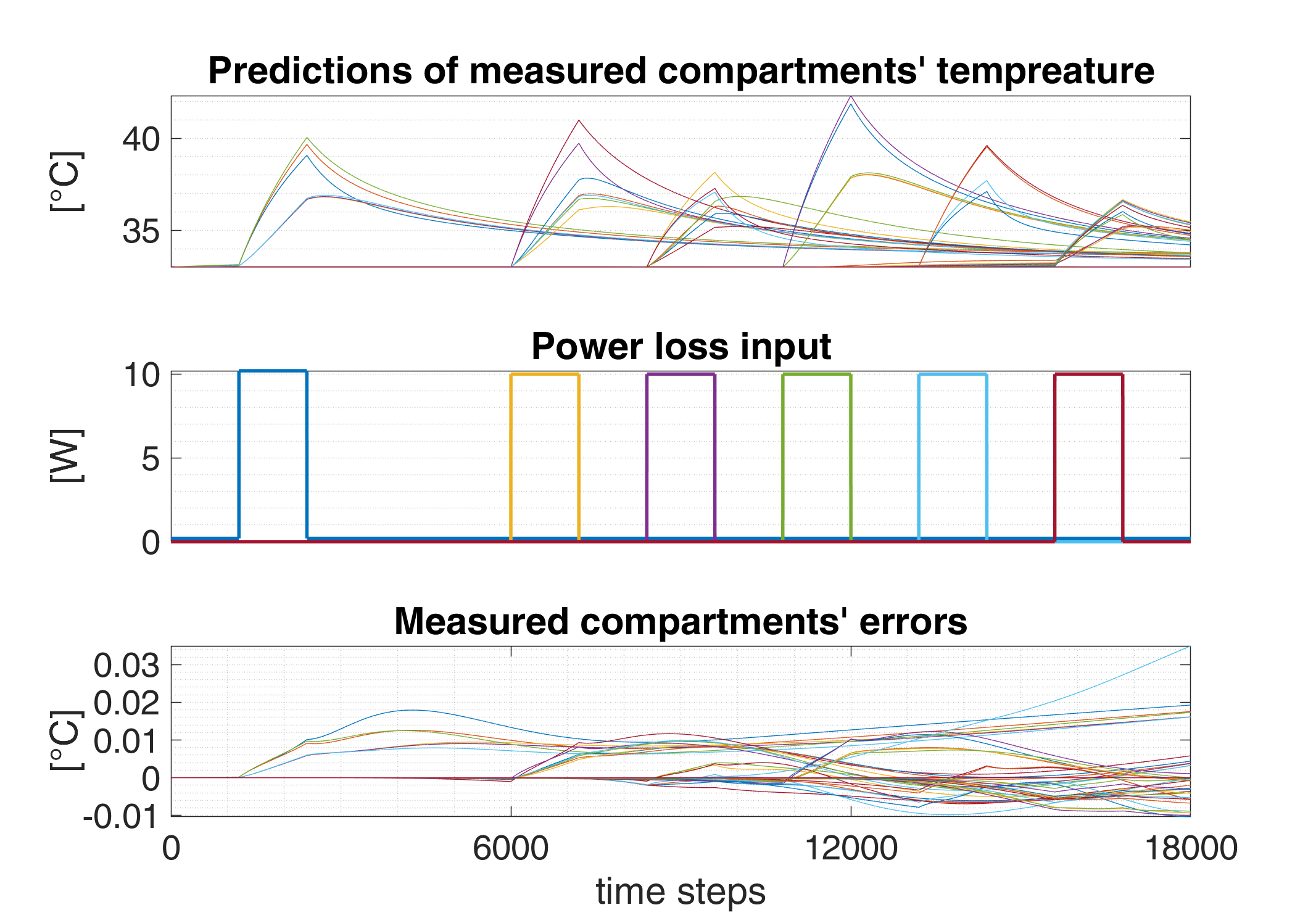}
\par\end{centering}
\caption{Prediction of temperature trends \textendash{} model generated with
weakly shared parametrization and identified using strongly shared
parametrization\label{fig:results_SpreadKSpreadData}}
\end{figure}

The convergence of parameters\,$\boldsymbol{k}$ during identification
process is depicted in Figure \ref{fig:Convergence-of-parameters}.
It can be seen that in the case of strongly shared parametrization,
all parameters converge to their true values (marked by crosses in
the graph). In the case of weakly shared parametrization, the EM algorithm
converges as well, but not to all true values of parameters\,$\boldsymbol{k}$.
It can be caused by lack of information about temperatures in unobserved
compartments (e.g. compartments representing diodes or rectifiers
in the true power module). Despite the EM algorithm not converging
to the true parameters in the weakly shared parametric model, the
trend of temperature predicted by the identified model stays valid
in some cases as can be seen in Figure \ref{fig:results_SpreadKSpreadData},
i.e. the identified model still explains measured temperatures relatively
correctly. This conclusion is probably valid if no specific temperature
fluctuation exists in unobserved compartments connected with remaining
parts of model with poorly identified connections (e.g. a connection
between rectifiers and Cu layer or a connection between diodes and
Cu layer).

The error in prediction of temperature depicted in Figure \ref{fig:results_SpreadKSpreadData}
is not greater than 4\% (maximum error of 0.3°C for temperature trend,
where the difference between ambient temperature and maximum temperature
is more than 9°C) for the synthetic model on relatively long-term
prediction (for prediction of 18000 time steps from the initial value
of temperature and with knowledge of the input vector $\boldsymbol{P}_{1:N}$
only). Note, that the same holds not only for the measured compartment,
but for all compartments in the model as well.

\subsection{Temperature prediction in dependency on covariance matrix structure
\label{sec:ComparisonCovarianceStructure}}

In this subsection we investigate the convergence properties of the
EM algorithm for identification of the proposed mesh-based compartment
thermal model in dependency on structure's constraint of the process
noise covariance matrix $Q$.

For the analysis of covariance matrix estimation, the weakly shared
parametrization is employed to generate data, while during identification
process strongly shared parametrization is assumed. It means that
the identified model is thus not identical to the ground truth, although
the structure (mesh-based discretization) of compartments is still
the same. In other words, using various parametrization for generating
data and for identifying model causes, that we do not know the true
form of auxiliary matrices $\mathcal{A},\mathcal{B}$ and $\mathcal{C}$
in equations \eqref{eq:Agivenbyk} and \eqref{eq:Bgivenbyz}. Moreover,
data are generated with process noise
\begin{equation}
\boldsymbol{w}_{t}\sim\mathcal{N}\left(\boldsymbol{w}_{t}|\boldsymbol{0},\sigma^{2}AA'\right),\label{eq:modelForComparisonQ}
\end{equation}
where $\sigma^{2}$ is set to value $10^{-4}$ and matrix $A$ is
defined by \eqref{eq:Agivenbyk}.

We investigate three kinds of parametrization of the covariance matrix
estimator similary as it is introduced in Section \ref{subsec:Structure-of-processCovariance}.
For covariance structure's constraint $Q_{\alpha LL'+\beta I}^{r}=\alpha^{r}LL'+\beta^{r}I_{n}$,
elements $l_{i,j}$ of matrix $L\in\mathbb{R}^{n\times n}$ are defined
as
\begin{eqnarray}
l_{i,j}=\left\{ \begin{array}{c}
1\\
0
\end{array}\right. & \begin{array}{l}
\textrm{\qquad if }(\mathcal{I}\mathrm{\text{diag}(\mathcal{C}\boldsymbol{1})\mathcal{J}'})_{i,j}\textrm{ is NOT }0\textrm{ and }j<n\\
\textrm{\qquad other}
\end{array} & \qquad i,j=1:n
\end{eqnarray}
where $\boldsymbol{1}$ is vector of all ones with the same dimension
as vector $\boldsymbol{k}$ and where notation from \eqref{eq:Agivenbyk}
is used.

\begin{figure}[h]
\begin{centering}
\begin{minipage}[t]{0.49\columnwidth}%
\begin{center}
\subfloat[State matrix $A$]{\includegraphics[width=0.9\textwidth]{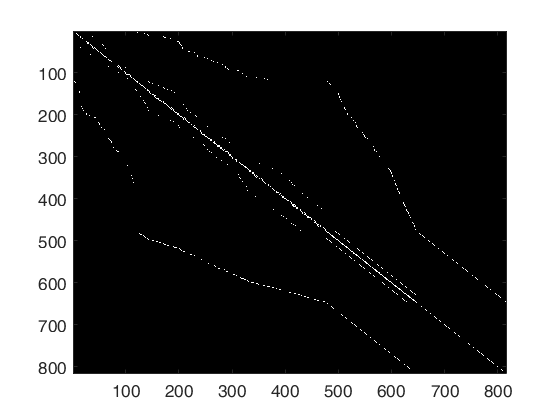}

}
\par\end{center}%
\end{minipage} %
\begin{minipage}[t]{0.49\columnwidth}%
\begin{center}
\subfloat[Matrix $LL'$ used for covariance matrix $Q$ estimation]{\includegraphics[width=0.9\textwidth]{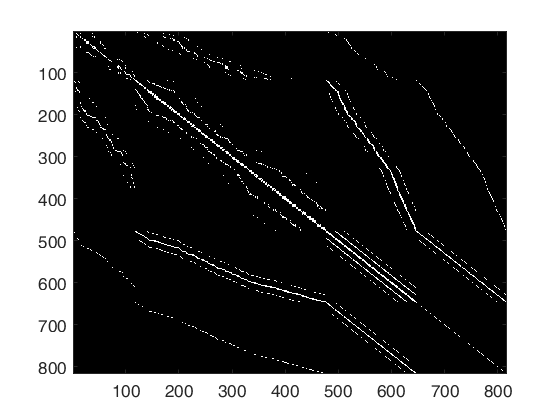}

}
\par\end{center}%
\end{minipage}
\par\end{centering}
\centering{}\caption{Illustration of matrix structure (white color corresponds to not zero
element, black color corresponds to zero element)}
\end{figure}

\begin{figure}[h]
\begin{centering}
\begin{minipage}[t]{0.33\columnwidth}%
\begin{center}
\subfloat[Constraint $Q\protect\overset{!}{=}qI_{n}.$]{\includegraphics[width=1\textwidth]{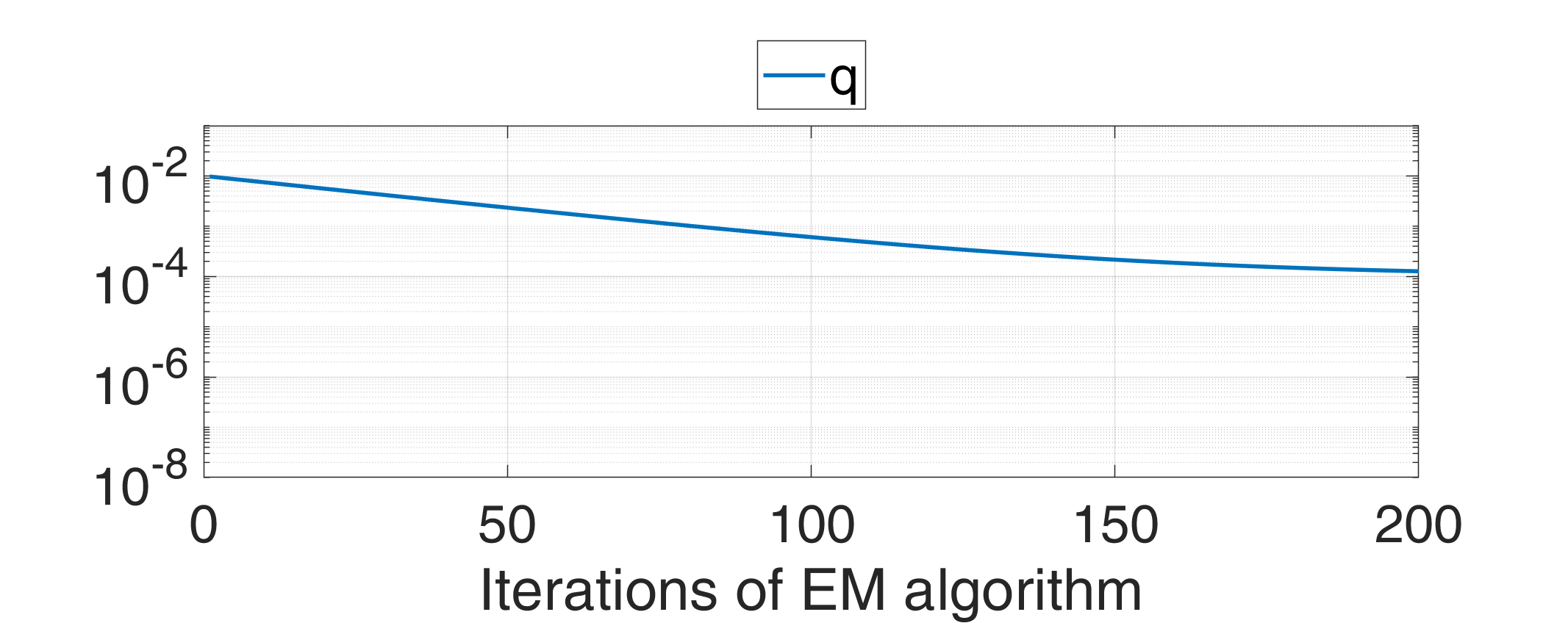}

}
\par\end{center}%
\end{minipage}%
\begin{minipage}[t]{0.33\columnwidth}%
\begin{center}
\subfloat[Constraint $Q\protect\overset{!}{=}\text{diag}\left(\boldsymbol{q}\right)$]{\includegraphics[width=1\textwidth]{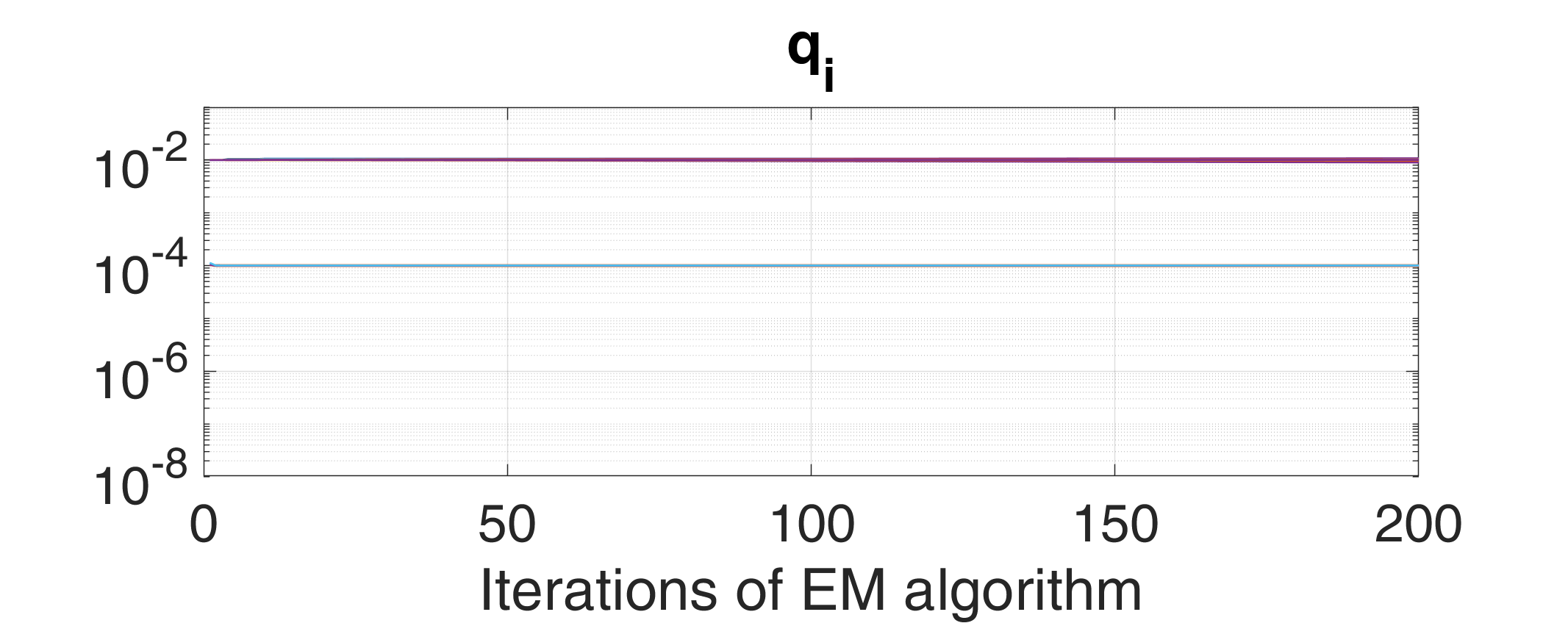}

}
\par\end{center}%
\end{minipage}%
\begin{minipage}[t]{0.33\columnwidth}%
\begin{center}
\subfloat[Constraint $Q\protect\overset{!}{=}\alpha LL'+\beta I_{n},$]{\includegraphics[width=1\textwidth]{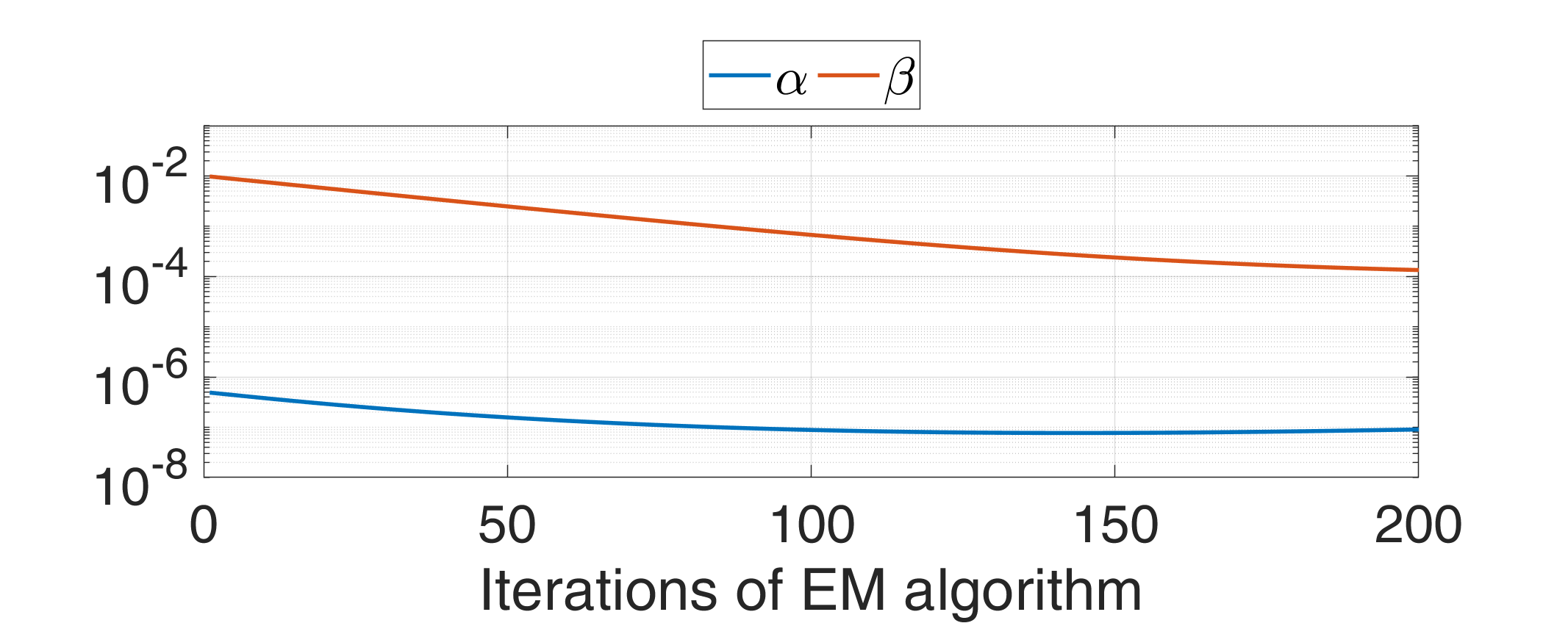}

}
\par\end{center}%
\end{minipage}
\par\end{centering}
\centering{}\caption{Convergence of process noise covariance matrix estimation \label{fig:Convergence-of-Q}}
\end{figure}

\begin{figure}[h]
\begin{centering}
\begin{minipage}[t]{0.33\columnwidth}%
\begin{center}
\subfloat[Constraint $Q\protect\overset{!}{=}qI_{n}.$]{\includegraphics[width=1\textwidth]{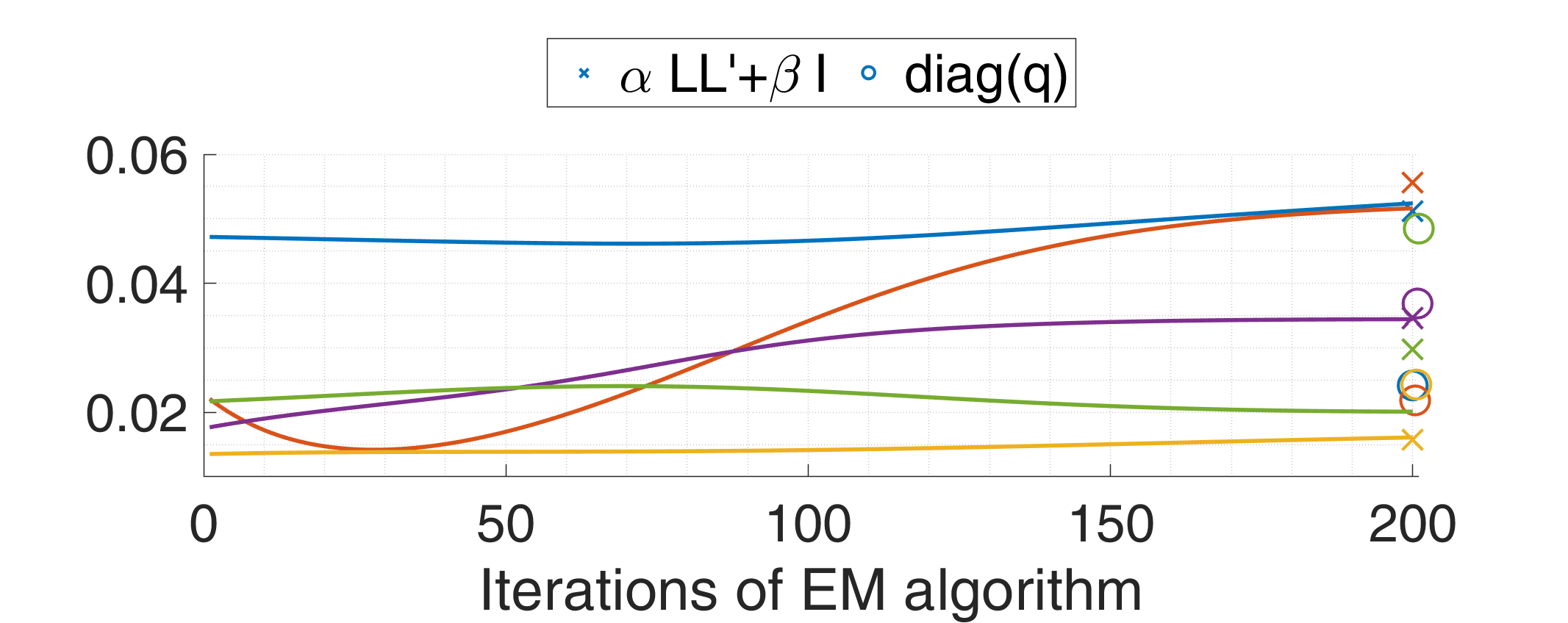}

}
\par\end{center}%
\end{minipage}%
\begin{minipage}[t]{0.33\columnwidth}%
\begin{center}
\subfloat[Constraint $Q\protect\overset{!}{=}\text{diag}\left(\boldsymbol{q}\right)$]{\includegraphics[width=1\textwidth]{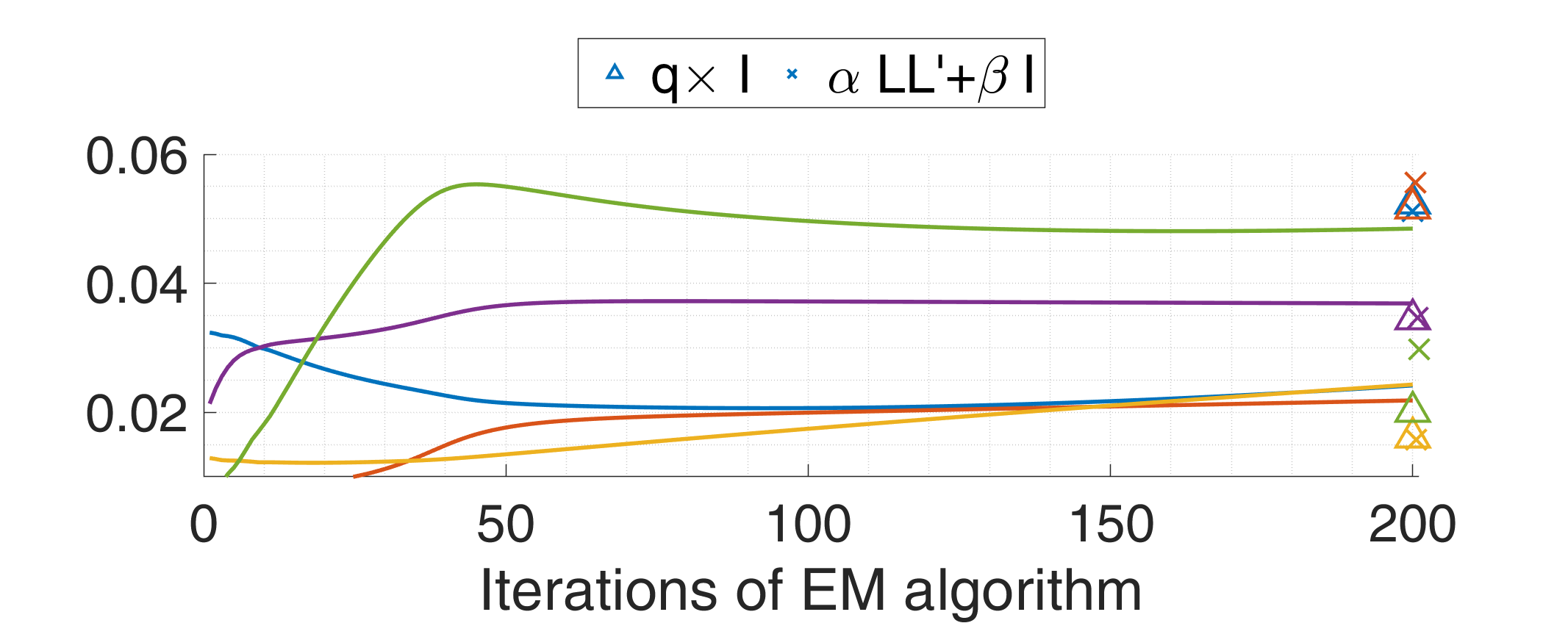}

}
\par\end{center}%
\end{minipage}%
\begin{minipage}[t]{0.33\columnwidth}%
\begin{center}
\subfloat[Constraint $Q\protect\overset{!}{=}\alpha LL'+\beta I_{n},$]{\includegraphics[width=1\textwidth]{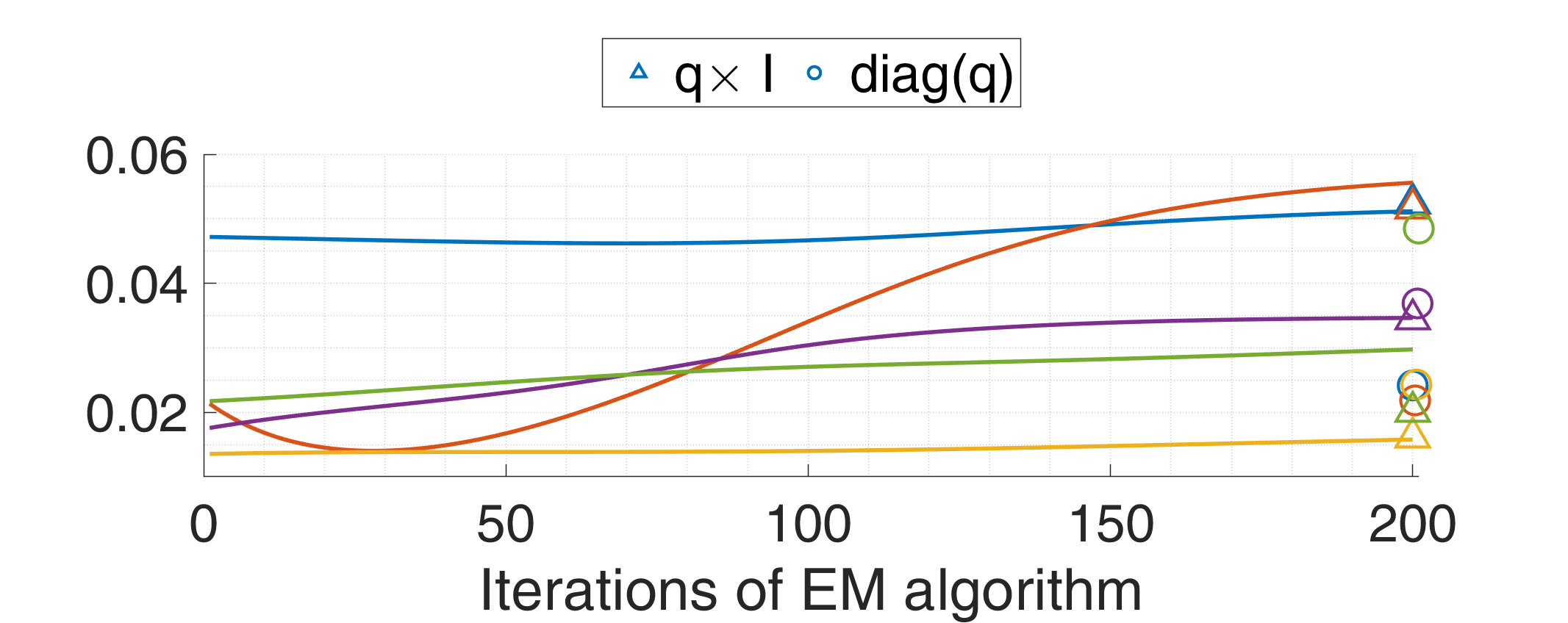}

}
\par\end{center}%
\end{minipage}
\par\end{centering}
\centering{}\caption{Convergence of values of parameters vector $\boldsymbol{k}$ in dependency
on the constraint of process noise covariance matrix \label{fig:Convergence-of-k-dependent-on-Q}}
\end{figure}

\begin{figure}[h]
\begin{centering}
\includegraphics[width=0.33\textwidth]{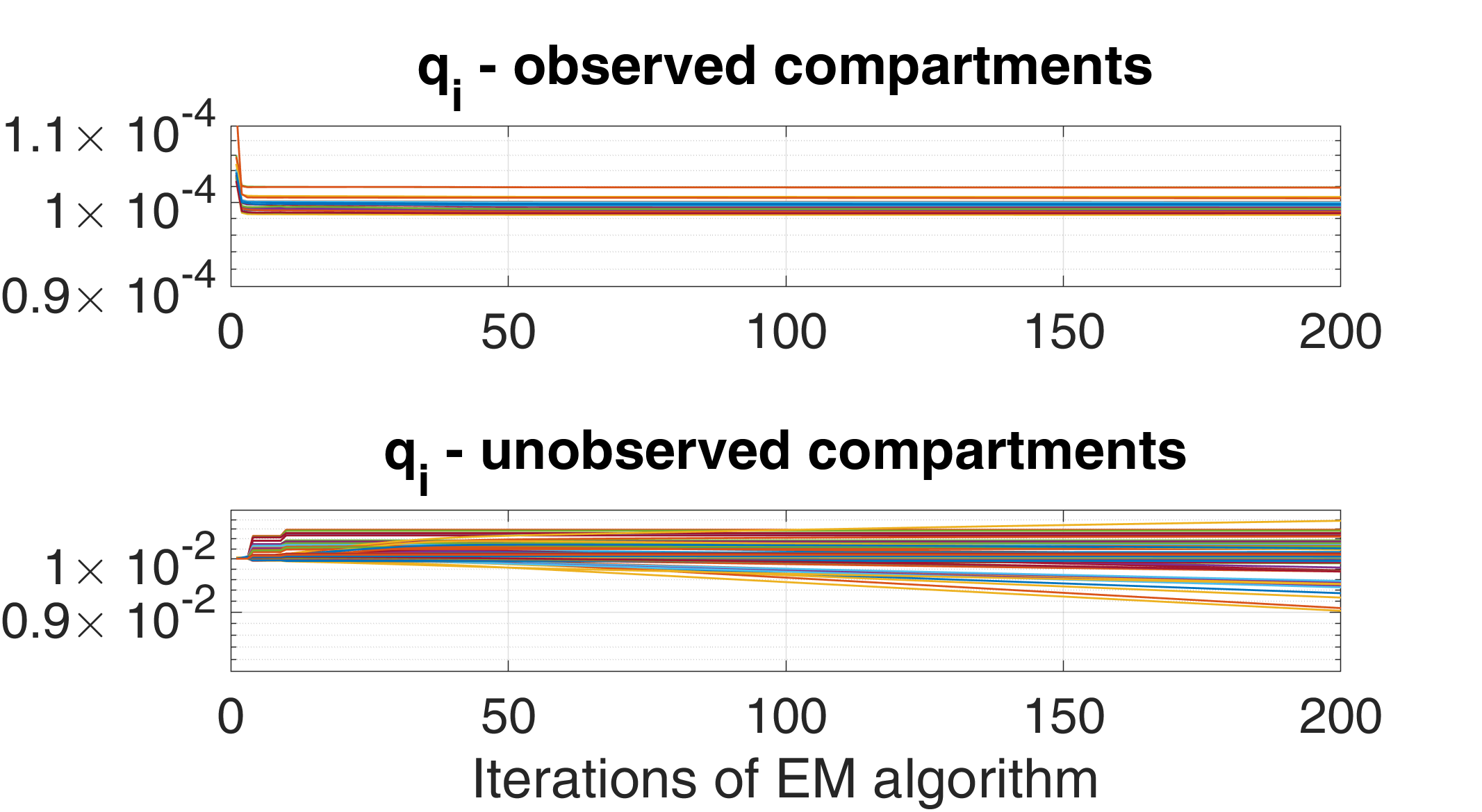}
\par\end{centering}
\centering{}\caption{Convergence of process noise covariance matrix estimation for the
constraint $Q\protect\overset{!}{=}\text{diag}\left(\boldsymbol{q}\right)$
\textendash{} separation of elements corresponding to observed compartments
and unobserved compartments\label{fig:Convergence-diag-observedVSunobserved}}
\end{figure}

The convergence of parameters describing constrained covariance matrix
$Q$ is depicted in Figure~\ref{fig:Convergence-of-Q}. The convergence
of values of parameters vector $\boldsymbol{k}$ dependent on the
choice of regularization of process noise covariance estimator is
depicted in Figure \ref{fig:Convergence-of-k-dependent-on-Q}. It
can be seen, that models using constraints on covariance matrix $Q$
in forms $Q\overset{!}{=}qI_{n}$ and $Q\overset{!}{=}\alpha LL'+\beta I_{n}$
give similar results and both of these form are sufficiently regularized.
Moreover with knowledge of true covariance matrix \eqref{eq:modelForComparisonQ}
and being aware of matrix $A$ is diagonally dominant, we can claim,
that these two approaches converge to the plausible values of covariance
parameters. From this point of view, the constraint $Q\overset{!}{=}\text{diag}\left(\boldsymbol{q}\right)$
seems to be overparameterized, since the convergence of selected elements
to the value $10^{-2}$, i.e. staying at the initial value, is not
well-founded. Figure \ref{fig:Convergence-diag-observedVSunobserved}
can give an explanation of this phenomenon. Diagonal elements of covariance
matrix taking higher values of variance ($10^{-2}$) are elements
just corresponding to unobserved compartments. Elements converging
to the true value of variance $10^{-4}$ are elements corresponding
to the observed compartments. Thus due to lack of information about
unobserved compartments, we are not able to identify the variance
correctly using constraint $Q\overset{!}{=}\text{diag}\left(\boldsymbol{q}\right)$.

\begin{figure}[H]
\begin{centering}
\begin{minipage}[t]{0.49\columnwidth}%
\begin{center}
\subfloat[Constraint $Q\protect\overset{!}{=}qI_{n}.$]{\includegraphics[width=1\textwidth]{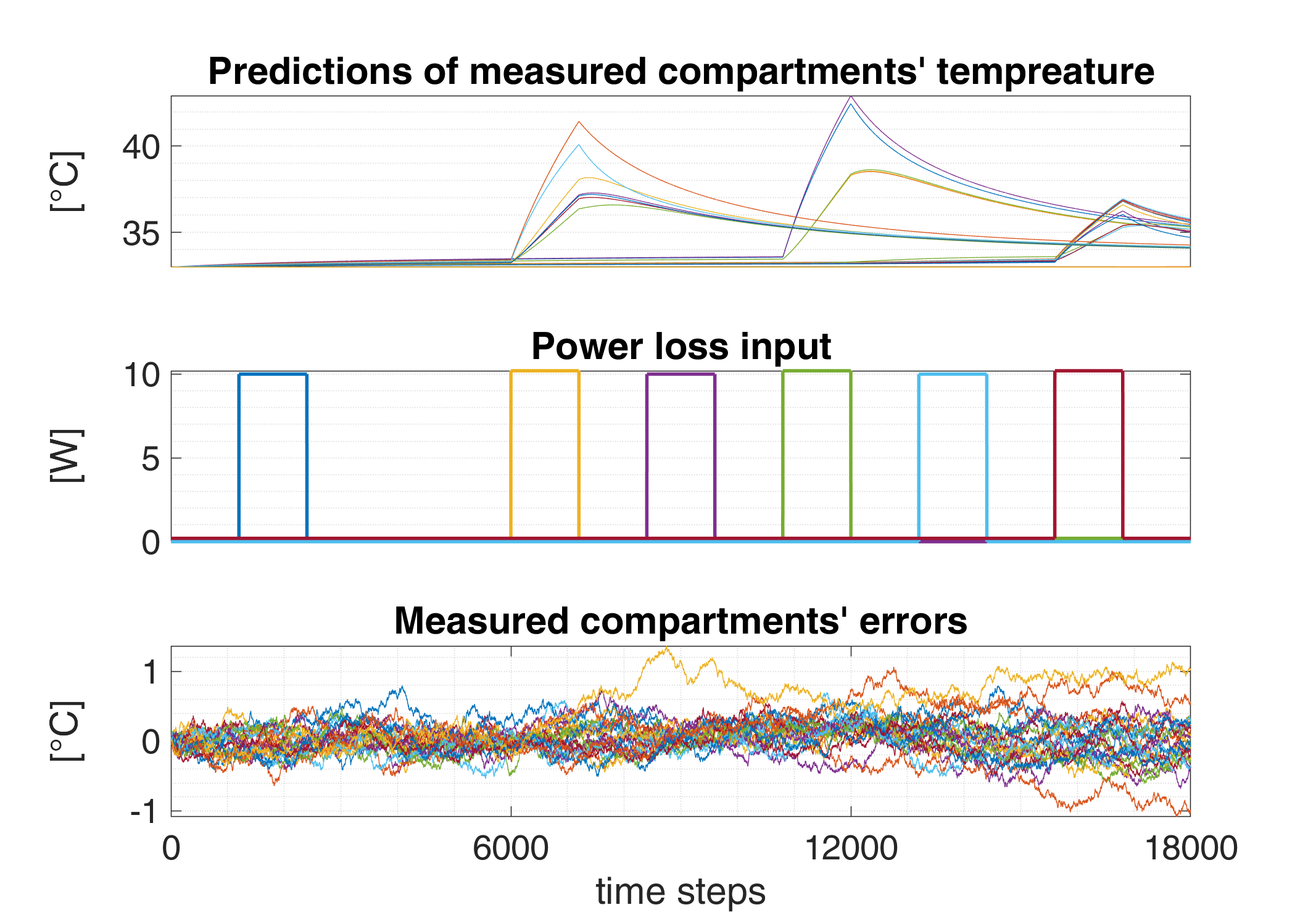}

}
\par\end{center}%
\end{minipage}%
\begin{minipage}[t]{0.49\columnwidth}%
\begin{center}
\subfloat[Constraint $Q\protect\overset{!}{=}\alpha LL'+\beta I_{n},$]{\includegraphics[width=1\textwidth]{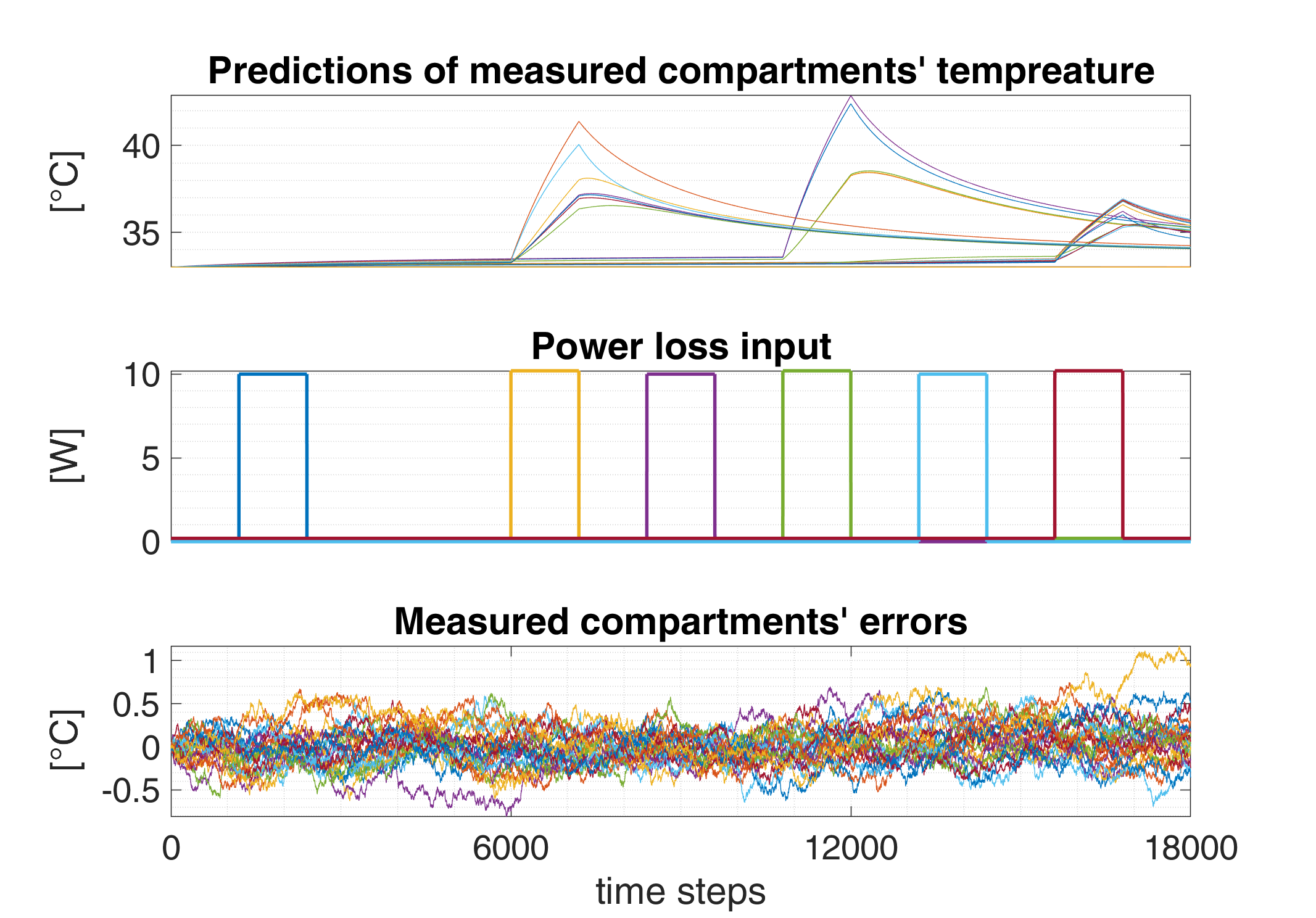}

}
\par\end{center}%
\end{minipage}
\par\end{centering}
\caption{Temperature prediction and its errors \label{fig:Temperature-prediction-and-it-errors}}
\end{figure}

The long-term predictions of 18000 time steps using models identified
with constrained covariance estimators $Q\overset{!}{=}qI_{n}$ and
$Q\overset{!}{=}\alpha LL'+\beta I_{n}$ respectively are depicted
in Figure \ref{fig:Temperature-prediction-and-it-errors}. Similar
as in subsection \ref{subsec:Convergence-of-parameters}, inputs of
prediction are the initial values of temperatures and vector $\boldsymbol{P}_{1:N}$
only. On the contrary, we do not know the true parameterization during
identification process (the parameterization used for identification
is different from the one used for data generation). Nevertheless,
the error of prediction is always smaller than 1 °C for temperature
trends, where the difference between ambient temperature and maximum
temperature is more than 10°C.

\section{Conclusion}

Mesh-based compartment thermal model and its identification procedure
using Expectation-Maximization algorithm was proposed. Using steady-state
covariance matrix in E-step of EM algorithm was suggested for speeding
up the identification algorithm and constraints on structure of process
noise covariance matrix in estimation procedure was investigated in
detail.

Preliminary tests on synthetic data indicated applicability of the
proposed thermal model and the identification approach. The selection
of parametrization has a strong impact on the possibility to identify
the model from incomplete temperature data. The strongly shared parameterized
models are better identifiable and furthermore enable to explain more
complicated models. However, the validation on real measured data
is needed and ought to be carried out by authors in the near future.

\ack{}{}

This research has been supported by the Ministry of Education, Youth
and Sports of the Czech Republic under the project OP VVV Electrical
Engineering Technologies with High-Level of Embedded Intelligence
CZ.02.1.01/0.0/0.0/18\_069/0009855 and by the UWB Student grant project
no. SGS-2018-009.

\appendix

\section{Expected terms for M-step\label{sec:Appendix_expected-terms-for-Mstep}}

Expected terms necessary for equations \eqref{eq:MLE_theta} and \eqref{eq:MLE_Q}
expressed using only statistics \eqref{eq:RTSSstatistics} obtained
from a previous E-step:

\begin{eqnarray*}
\sum\limits _{t=1}^{N-1}E_{\Phi^{r}(T)}\left\{ \Delta\boldsymbol{T}_{t}\Delta\boldsymbol{T}_{t}'\right\} =\frac{1}{\Delta\tau^{2}}\left(XX'-XZ'-(XZ')'+ZZ'\right)\\
\sum\limits _{t=1}^{N-1}E_{\Phi^{r}(T)}\left\{ M_{t}'Q^{-1}M_{t}\right\} =\\
\hspace{1.3cm}\text{=}\left[\begin{array}{cc}
\mathcal{C}'\left(\left(\mathcal{I}'Q^{-1}\mathcal{I}\right)\circ\left(\mathcal{J}'XX'\text{\ensuremath{\mathcal{J}}}\right)\right)\mathcal{C} & -\mathcal{C}'\left(\left(\mathcal{I}'Q^{-1}\mathcal{B}\right)\circ\left(\mathcal{J}'XU'\right)\right)\mathcal{A}\\
-\mathcal{A}'\left(\left(\mathcal{B}'Q^{-1}\mathcal{I}\right)\circ\left(\left(XU'\right)'\mathcal{J}\right)\right)\mathcal{C} & \mathcal{A}'\left(\mathcal{B}'Q^{-1}\mathcal{B}\circ UU'\right)\mathcal{A}
\end{array}\right]\\
\sum\limits _{t=1}^{N-1}E_{\Phi^{r}(T)}\left\{ M_{t}\boldsymbol{\theta}\boldsymbol{\theta}'M_{t}'\right\} =\left\{ \boldsymbol{\theta}\boldsymbol{\theta}'\equiv\left[\begin{array}{cc}
\boldsymbol{kk}' & \boldsymbol{kz}'\\
\boldsymbol{zk}' & \boldsymbol{zz}'
\end{array}\right]\right\} =\\
\hspace{1.3cm}=\mathcal{I}\left(\left(\mathcal{C}\boldsymbol{kk}'\mathcal{C}'\right)\circ\left(\mathcal{J}'XX'\text{\ensuremath{\mathcal{J}}}\right)\right)\mathcal{I}'-\text{\ensuremath{\mathcal{B}}}\left(\left(\mathcal{A}\boldsymbol{zk}'\mathcal{C}'\right)\circ\left(\left(XU'\right)'\mathcal{J}\right)\right)\mathcal{I}'+\\
\hspace{1.3cm}\hphantom{=}-\mathcal{I}\left(\left(\mathcal{C}\boldsymbol{kz}'\mathcal{A}'\right)\circ\left(\mathcal{J}'XU'\right)\right)\mathcal{B}'+\text{\ensuremath{\mathcal{B}}}\left(\left(\mathcal{A}\boldsymbol{zz}'\mathcal{A}'\right)\circ UU'\right)\mathcal{B}'\\
\sum\limits _{t=1}^{N-1}E_{\Phi^{r}(T)}\left\{ M_{t}Q^{-1}\Delta\boldsymbol{T}_{t}\right\} =\Delta t^{-1}\left[\begin{array}{c}
-\mathcal{C}^{T}\text{diag}\left(\mathcal{J}^{T}\left(XZ'-XX'\right)\left(Q^{-1}\right)'\mathcal{I}\right)\\
\mathcal{A}^{T}\text{diag}\left(\left(ZU'-XU'\right)'\left(Q^{-1}\right)'\mathcal{B}\right)
\end{array}\right]\\
\sum\limits _{t=1}^{N-1}E_{\Phi^{r}(T)}\left\{ \Delta\boldsymbol{T}_{t}\boldsymbol{\theta}'M_{t}'\right\} =\left\{ \boldsymbol{\theta}\equiv\left[\begin{array}{c}
\boldsymbol{k}\\
\boldsymbol{z}
\end{array}\right]\right\} =\\
\hspace{1.3cm}=\Delta t^{-1}\left(XZ'-XX'\right)'\mathcal{J}\text{diag}(-C\boldsymbol{k})\mathcal{I}^{T}+\Delta t^{-1}\left(ZU'-XU'\right)\text{diag}(\mathcal{A}\boldsymbol{z})\mathcal{B}^{T},
\end{eqnarray*}
where symbol $\circ$ stands for Hadamard product, i.e. element-wise
multiplication, and operator $\text{diag}(\cdot)$ applied on a vector
creates diagonal matrix with the vector values on the main diagonal
and operator $\text{diag}(\cdot)$ applied on a matrix extracts the
main diagonal and the rest of elements replaces with zeros.

\bibliographystyle{unsrt}
\addcontentsline{toc}{section}{\refname}\bibliography{usedBibliography}

\end{document}